\documentclass[12pt,dvips]{article}

\usepackage{rotating}
\usepackage{epsfig}
\usepackage{amssymb}
\usepackage{color}

\def\lsim{\mathrel{\raise.3ex\hbox{$<$\kern-.75em\lower1ex\hbox{$\sim$}}}}
\def\gsim{\mathrel{\raise.3ex\hbox{$>$\kern-.75em\lower1ex\hbox{$\sim$}}}}

\begin{document}

\begin{titlepage}

\begin{flushright}
KAIST-TH 2007/02 \\[-0.1cm]
hep-ph/0703163\\[1mm]
\today
\end{flushright}

\vskip 1.5cm

\begin{center}
{\Large\bf LHC Signature of Mirage Mediation}\\[1.5cm]
{W.S. Cho$^1$, Y.G. Kim$^{1,2}$, K.Y. Lee,$^1$, C.B. Park$^1$ 
         and Y. Shimizu$^1$}
\end{center}

\vskip 0.5cm

{\small
\begin{center}
$^1$ {\it Department of Physics, KAIST, Daejon 305--017, Korea} \\
$^2$ {\it ARCSEC, Sejong University, Seoul 143-747, Korea}\\
\end{center}
}

\vskip 4.cm
\begin{abstract}
\noindent
We study LHC phenomenology of mirage mediation scenario in which 
anomaly and modulus contributions to soft SUSY breaking terms are comparable
to each other. A Monte Carlo study of mirage mediation, with model parameters $\alpha=1$,$~M_0=500$ GeV,
$n_M=1/2$, $n_H=1$ and $\rm{tan}\beta=10$, is presented. It is shown that 
masses of supersymmetric particles can be measured in a model independent way, 
providing information on SUSY breaking sector. In particular, the mass ratio of gluino to
the lightest neutralino for the benchmark scenario 
is determined to be $1.9 \lesssim m_{\tilde g}/m_{\tilde\chi_1^0} \lesssim 3.1$,
well reproducing theoretical input value of $m_{\tilde g}/m_{\tilde\chi_1^0} \simeq 2.5$
which is quite distinctive from the predictions $m_{\tilde g}/m_{\tilde\chi_1^0} \gtrsim 6$
of other SUSY scenarios in which gaugino masses are unified at the GUT scale.
The model parameters of mirage mediation can be also determined from various kinematic distributions. 
\end{abstract}
\end{titlepage}

\section{Introduction}
\label{sec:introduction}

Weak scale supersymmetry (SUSY) is one of the most promising candidates
of new physics beyond the standard model (SM) \cite{MSSM}.
It provides a solution for gauge hierarchy problem and complies with gauge coupling unification.
Another nice feature of the SUSY theory with R-parity conservation is that 
the lightest supersymmetric particle (LSP) is a natural candidate for 
the non-baryonic dark matter (DM) in the universe. 

SUSY phenomenology crucially depends on the masses of SUSY particles and their properties, 
which might result from spontaneous SUSY breaking in a hidden sector. 
The SUSY breaking is communicated to visible sector 
through some messenger interactions depending on models.
Soft SUSY breaking terms of visible matter fields are then determined 
such that different SUSY breaking and mediating mechanism leads to 
different pattern of masses and properties of SUSY particles.

Recently Kachru $et~ al.$ (KKLT) has provided a concrete set-up of string compactification,
in which all moduli are fixed and Minkowski (or de Sitter) vacuum is achieved \cite{kklt}.
The KKLT-type moduli stabilization scenario leads to an interesting pattern of soft 
SUSY breaking terms, to which modulus and anomaly contributions are comparable to each other
if gravitino mass $m_{3/2}\sim 10$ TeV \cite{choi1}. 
A noticeable feature of the mixed modulus-anomaly mediation ($a.k.a$ mirage mediation)
is that soft masses are unified 
at a mirage messenger scale \cite{Choi:2005uz}
\begin{eqnarray}
M_{mir} = M_{GUT} \left(\frac{m_{3/2}}{M_{Pl}}\right)^{\alpha/2}
\end{eqnarray}
with $\alpha$ representing the ratio of the anomaly to modulus mediation.
The mirage messenger scale $M_{mir}$ is hierarchically lower than $M_{GUT}$ 
for a positive $\alpha\sim O(1)$. 
Such a low (mirage) unification scale for soft masses leads to a SUSY mass spectrum which is quite
distinctive from those in other SUSY breaking scenarios such as mSUGRA, gauge mediation
and anomaly mediation.  
Some phenomenological aspects of the mirage mediation have been investigated by
several authors \cite{mirage3,cklos}.

In this paper, we investigate the LHC signatures of the mirage mediation
performing a Monte Carlo study for a benchmark point in the scenario and show that
SUSY particle masses can be determined in a model independent way, providing some 
valuable information on SUSY breaking sector.
If the weak scale SUSY is realized in nature, SUSY particles would be produced copiously
at the LHC \cite{LHC, beenakker}, which is scheduled to start in 2007. 
Gluinos and squarks, which are directly produced from the proton-proton collision, 
will decay to the LSP and SM particles in the end, 
with non-colored SUSY particles as intermediate states
in general. The precise measurement of the masses of SUSY particles
might be possible with reconstruction of the cascade decay chains \cite{Bachacou:1999zb,weiglein}.
Kinematic edges and thresholds of various invariant mass distributions 
can be measured experimentally
and then the SUSY particle masses would be determined in a model independent way.
In turn, SUSY breaking mechanism might be reconstructed 
from the measured SUSY spectrum and signatures.

This paper is organized as follows.
The basic features of the mirage mediation are briefly described in section 2.
The Monte Carlo study for a benchmark point is presented in section 3.
Section 4 is devoted to the conclusions.

\section{Mirage Mediation}
\label{sec:mirage}
In KKLT-type moduli stabilization scenario, the light modulus $T$ which determines
the SM gauge couplings is stabilized by non-perturbative effects
and the SUSY-breaking source is sequestered from the visible sector.
The non-perturbative stabilization of $T$ by modulus superpotential 
results in a suppression of the modulus $F$ component: 
\begin{eqnarray}
\frac{F^T}{T} \sim \frac{m_{3/2}}{{\rm ln}(M_{Pl}/m_{3/2})},
\end{eqnarray}
which is comparable to
anomaly mediated soft mass of $O(m_{3/2}/4\pi^2)$ for $m_{3/2}$ near the TeV scale.
The soft terms of visible fields are then determined by the modulus mediation
and the anomaly mediation if the SUSY breaking brane is sequestered from the visible sector.
The soft terms of canonically normalized visible fields are given by
\begin{eqnarray}
{\cal L}_{\rm
soft}&=&-\frac{1}{2}M_a\lambda^a\lambda^a-\frac{1}{2}m_i^2|\phi_i|^2
-\frac{1}{6}A_{ijk}y_{ijk}\phi_i\phi_j\phi_k+{\rm h.c.},
\end{eqnarray}
where $\lambda^a$ are gauginos, $\phi_i$ are the scalar component
of visible matter superfields $\Phi_i$ 
and $y_{ijk}$ are the canonically normalized Yukawa couplings. 
For $F^T/T\sim m_{3/2}/4\pi^2$,
the soft parameters  at energy scale just below  $M_{GUT}$ are
determined by the modulus-mediated and anomaly-mediated contributions 
which are comparable to each other.
One then finds the boundary values of gaugino masses, trilinear couplings
and sfermion masses at $M_{GUT}$ are given by
\begin{eqnarray}
M_a&=& M_0 \Big[\,1+\frac{\ln(M_{Pl}/m_{3/2})}{16\pi^2} b_a
g_a^2\alpha\,\Big],\nonumber \\
A_{ijk}&=&M_0\Big[\,(a_i+a_j+a_k)
-\frac{\ln(M_{Pl}/m_{3/2})}{16\pi^2}(\gamma_i+\gamma_j+\gamma_k)\alpha\,\Big],
\nonumber \\
m_i^2&=&M_0^2\Big[\,c_i-\,\frac{\ln(M_{Pl}/m_{3/2})}{16\pi^2}
\theta_i\alpha-\left(\frac{\ln(M_{Pl}/m_{3/2})}{16\pi^2}\right)^2\dot{\gamma}_i\alpha^2\,\Big],
\label{eq:bc1}
\end{eqnarray}
where $\alpha$ represents the anomaly to modulus mediation ratio:
\begin{eqnarray}
\alpha \equiv \frac{m_{3/2}}{M_0~{\rm ln}(M_{\rm Pl}/m_{3/2})}~,
\end{eqnarray} with $M_0$ the pure modulus mediated gaugino mass, while
$a_i$ and $c_i$ parameterize the pattern of the pure modulus mediated soft masses. 
The one-loop beta function coefficient $b_a$, the anomalous dimension $\gamma_i$
and its derivative ${\dot\gamma_i}$ and $\theta_i$ are defined by
\begin{eqnarray}
b_a&=&-3{\rm tr}\left(T_a^2({\rm Adj})\right)
        +\sum_i {\rm tr}\left(T^2_a(\phi_i)\right),
\nonumber \\
\gamma_i&=&2\sum_a g^2_a C^a_2(\phi_i)-\frac{1}{2}\sum_{jk}|y_{ijk}|^2,
\nonumber \\
\dot{\gamma}_i&=&8\pi^2\frac{d\gamma_i}{d\ln\mu},\nonumber \\
\theta_i&=&4\sum_a g^2_a C^a_2(\phi_i)-\sum_{jk}|y_{ijk}|^2(a_i+a_j+a_k) ,
\end{eqnarray} where the
quadratic Casimir $C^a_2(\phi_i)=(N^2-1)/2N$ for a fundamental
representation $\phi_i$ of the gauge group $SU(N)$,
$C_2^a(\phi_i)=q_i^2$ for the $U(1)$ charge $q_i$ of $\phi_i$, and
$\omega_{ij}=\sum_{kl}y_{ikl}y^*_{jkl}$ is assumed to be diagonal.
The explicit expressions of $b_a$, $\gamma_i$, $\dot{\gamma_i}$ and $\theta$ in
the MSSM are given in appendix A.
In this prescription, generic mirage mediation is parameterized by
\begin{eqnarray} 
\alpha, \,\, M_0,\,\, a_i,\,\, c_i,\,\, \tan\beta,
\end{eqnarray} where
${\rm tan}\beta$ is the ratio of the vacuum expectation values of the two neutral Higgs fields.

An interesting feature called mirage mediation 
arises from the soft masses of Eq. (\ref{eq:bc1}) at $M_{GUT}$, due to the correlation 
between the anomaly mediation and the RG evolution of soft parameters.
The low energy gaugino masses are given by \cite{Choi:2005uz}

\begin{eqnarray}
\label{lowgaugino} M_a(\mu)=M_0\left[\,
1-\frac{1}{8\pi^2}b_ag_a^2(\mu)\ln\left(\frac{M_{\rm mir}}{\mu}\right)\,\right] 
=\frac{g_a^2(\mu)}{g_a^2(M_{\rm mir})}M_0, 
\label{gauginomass}
\end{eqnarray} implying that the gaugino masses are unified at
$M_{\rm mir}$, while the gauge couplings are unified at $M_{GUT}$.
If the $y_{ijk}$ is small or $a_i+a_j+a_k = c_i+c_j+c_k=1$,
the low energy values of $A_{ijk}$ and $m_i^2$ are given by
\cite{Choi:2005uz}
\begin{eqnarray}
\label{lowsfermion}
A_{ijk}(\mu)&=& M_0\left[\,a_i+a_j+a_k+
\frac{1}{8\pi^2}(\gamma_i(\mu)
+\gamma_j(\mu)+\gamma_k(\mu))\ln\left(\frac{M_{\rm
mir}}{\mu}\right)\,\right], \nonumber \\
m_i^2(\mu)&=&M_0^2\left[\,c_i-\frac{1}{8\pi^2}Y_i\left(
\sum_jc_jY_j\right)g^2_Y(\mu)\ln\left(\frac{M_{GUT}}{\mu}\right)\right.
\nonumber \\
&+&\left.\frac{1}{4\pi^2}\left\{
\gamma_i(\mu)-\frac{1}{2}\frac{d\gamma_i(\mu)}{d\ln\mu}\ln\left(
\frac{M_{\rm mir}}{\mu}\right)\right\}\ln\left( \frac{M_{\rm
mir}}{\mu}\right)\,\right], 
\label{scalarmass}
\end{eqnarray}
where $Y_i$ is the $U(1)_Y$
charge of $\phi_i$. 
Therefore, the first- and second-generation sfermion masses are also unified 
at the $M_{mir}$ scale if the modulus-mediated squark and
slepton masses have a common value, i.e.  $c_{\tilde{q}}=c_{\tilde{\ell}}~ (\equiv c_M)$.

Phenomenology of mirage mediation is quite sensitive to 
the anomaly to modulus mediation ratio $\alpha$ as well as the parameters $a_i$ and $c_i$
\cite{Choi:2005uz,mirage3,cklos}.
When $\alpha$ increases from zero to a positive value of order unity,
the nature of the neutralino LSP is changed from bino-like to Higgsino-like via 
a bino-Higgsino mixing region. This feature can be understood from
the dependence of the gaugino masses on $\alpha$. The gluino mass $M_3$ decreases as
$\alpha$ increases, while the bino mass $M_1$ increases. 
Smaller $M_3$ then leads to smaller $|m_{H_u}^2|$ and Higgsino mass parameter $|\mu|$ 
at the weak scale through smaller stop mass square.
While the gaugino masses are not sensitive to $a_i$ and $c_i$, 
the low energy squark, slepton and Higgs masses depend on those parameters through
their boundary values at $M_{GUT}$ and their RG evolutions and 
the mass mixing induced by the low energy $A$ parameters.

For the original KKLT compactification of type IIB string theory \cite{kklt}, one finds that
$\alpha=1,~a_i=c_i=1-n_i$, where $n_i$ are modular weights of 
the visible sector matter fields depending on the origin of the matter fields.
The corresponding mirage messenger scale for $\alpha=1$ is given by $M_{mir} \sim 3\times 10^9$ GeV.
Such an intermediate (mirage) unification scale which is hierarchically smaller than $M_{GUT}$, 
leads to quite a degenerated sparticle mass spectrum at EW scale, 
compared to that for mSUGRA-type pure modulus mediation ($\alpha=0$).
In this paper, we will consider the intermediate scale mirage mediation ($\alpha=1$) 
as a benchmark scenario for a detailed Monte Carlo study of LHC phenomenology. 

\section{Collider signatures}
\label{sec:collider}

\subsection{A benchmark point}

We perform a Monte Carlo study for LHC signatures of mirage mediation 
with the following model parameters;
\begin{eqnarray}
\alpha=1,\,\,M_0=500~{\rm GeV},\,\,a_M=c_M=1/2,\,\,a_H=c_H=0,\,\,{\rm tan}\beta=10, 
\label{modelparameters}
\end{eqnarray}
where $c_M$ is a common parameter which parameterize the pattern of
the pure modulus mediated masses for squarks and sleptons and $c_H$ for soft Higgses. 
This choice of model parameters is denoted as a blue dot on the $(\alpha,M_0)$ plane
of Fig.1, which was taken from the figure 12 (c) in ref.\cite{cklos}.
In the Fig.1, the magenta stripe corresponds to the parameter region 
giving a thermal relic density consistent with the recent WMAP observation \cite{wmap},
$i.e.$ $0.085<\Omega_{DM} h^2 < 0.119$. The benchmark point is also consistent
with constraints on particle spectra and $b\rightarrow s\gamma$ branching ratio. 

\begin{figure}[ht!]
\vskip 0.6cm
\begin{center}
\epsfig{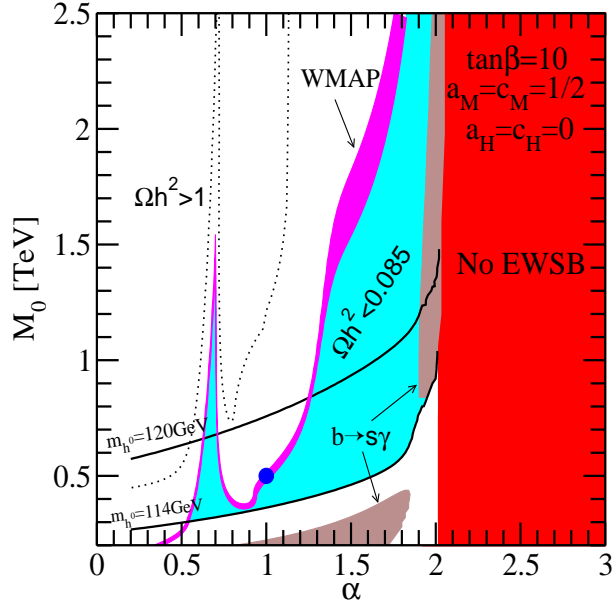}
\end{center}
\vskip 0.4cm
\caption{\it Parameter space $(\alpha, M_0)$ with $a_H=c_H=0$, $a_M=c_M=1/2$ and ${\rm tan\beta}=10$.
The blue point corresponds to our benchmark point, which satisfy relic density bound of WMAP observation
and is consistent with other experimental constraints on particle masses and 
$b\rightarrow s\gamma$ branching ratio.}
\label{fig:parameters}
\end{figure}

The SUSY particle mass spectrum at the electroweak(EW) scale was computed by solving 
the RG equations with the model parameter set (\ref{modelparameters}). 
For the gluino and the first two generation squark masses, we find 
\begin{eqnarray}
m_{\tilde g}=884.4~ {\rm GeV},~ m_{\tilde d_L (\tilde u_L)}=776.0~ (771.9)~{\rm GeV},
~m_{\tilde d_R (\tilde u_R)}=733.5~ (741.8)~{\rm GeV},
\label{mass1}
\end{eqnarray}
while the masses of the third generation squarks are
\begin{eqnarray}
m_{\tilde b_{1(2)}} = 703.9~(734.6)~{\rm GeV},
~m_{\tilde t_{1(2)}} = 545.3~(782.0)~{\rm GeV}
\end{eqnarray}
On the other hand, the slepton and sneutrino masses are
\begin{eqnarray}
m_{\tilde e_{R(L)}} = 382.0~(431.5)~{\rm GeV}, 
~m_{\tilde\tau_{1(2)}}= 378.9~(435.4)~{\rm GeV},
~m_{\tilde\nu_L} = 424.1~{\rm GeV}.
\end{eqnarray}
For the neutralino and chargino masses, we have
\begin{eqnarray}
m_{\tilde \chi^0_{1,2,3,4}}= \{355.1,~ 416.1,~ 478.7,~ 535.6\}~{\rm GeV},
~m_{\tilde \chi^\pm_{1,2}}=\{408.2,~533.5\}~{\rm GeV},
\end{eqnarray}
which correspond to the following bino,wino and Higgsino masses at electroweak scale,
\begin{eqnarray}
M_1=367~ {\rm GeV},~M_2=461~{\rm GeV},~\mu=475~{\rm GeV} ~~({\rm for~tan\beta=10}).
\end{eqnarray}
Finally, Higgs masses are given by
\begin{eqnarray}
m_h = 115~ {\rm GeV}, m_H=528.6~ {\rm GeV}, m_A=528.3~ {\rm GeV}, m_{H^\pm} =534.4~ {\rm GeV}.
\end{eqnarray}

\begin{figure}[ht!]
\vskip 0.6cm
\begin{center}
\epsfig{figure=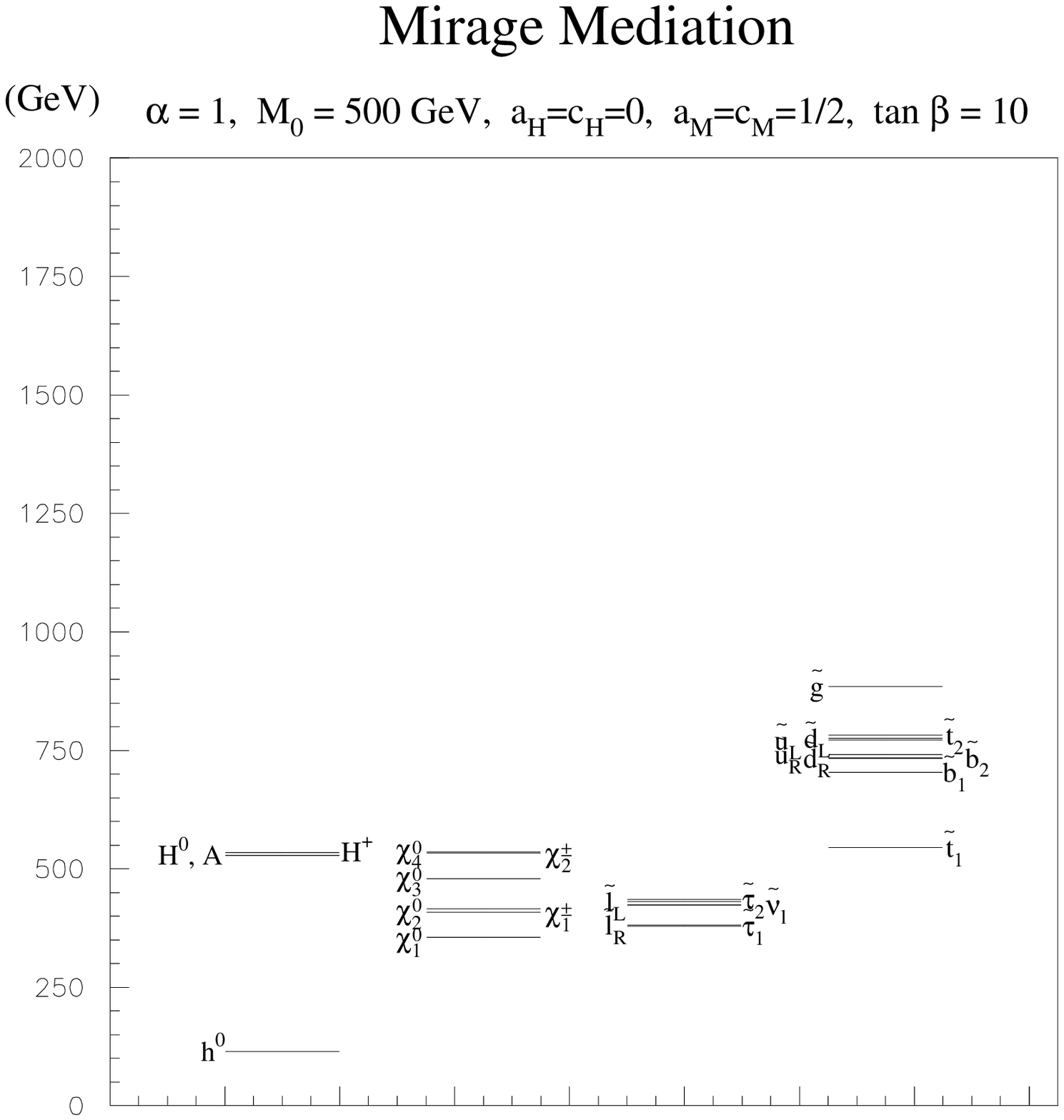,width=7cm,height=8cm}
\epsfig{figure=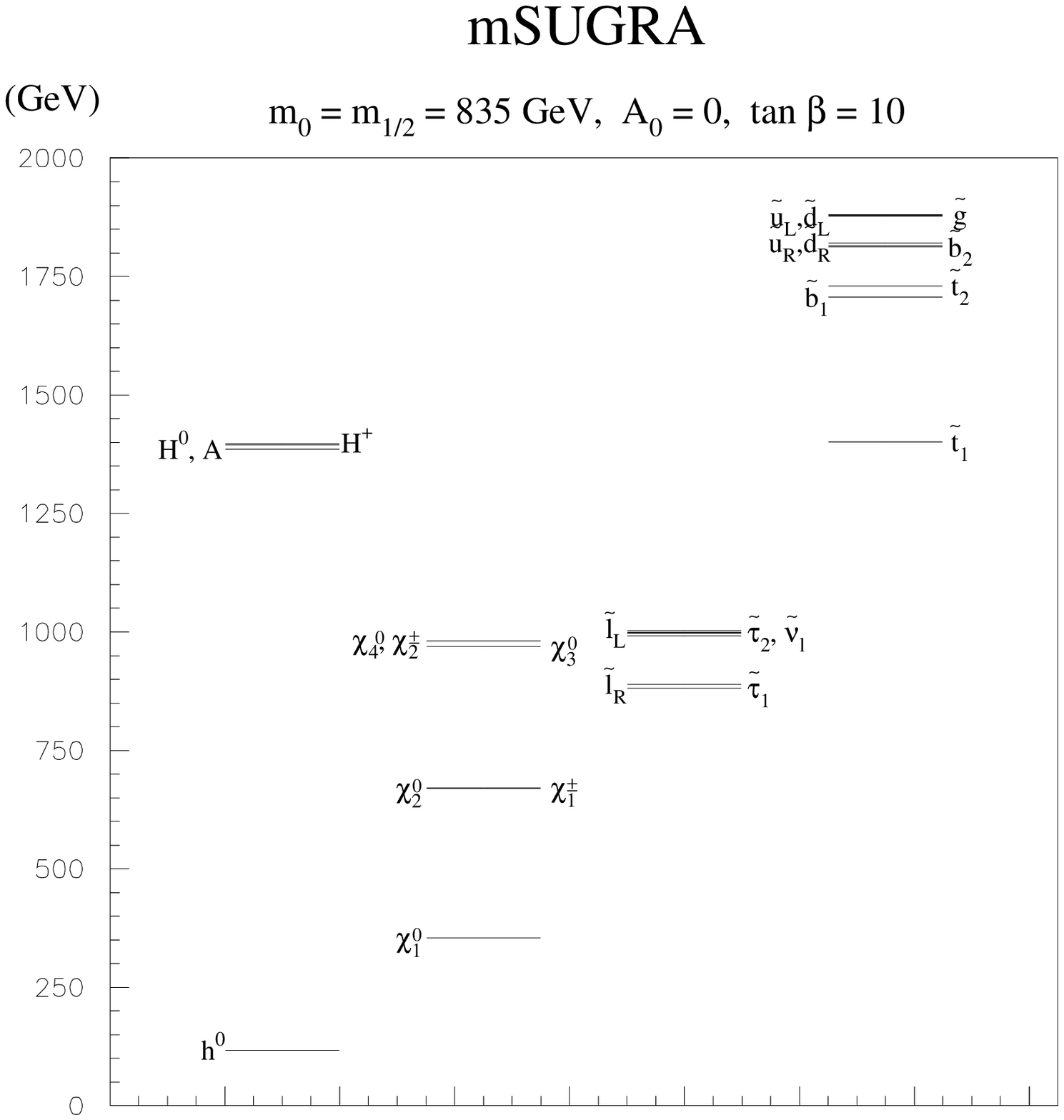,width=7cm,height=8cm}
\end{center}
\vskip 0.4cm
\caption{\it The mass spectrum for (a) the benchmark point in mirage mediation 
and (b) a mSUGRA point. Both cases give the same mass for the lightest neutralino.}
\label{fig:masses}
\end{figure}

Fig. \ref{fig:masses}(a) shows the mass spectrum for the benchmark point. 
For comparison, we also show the spectrum of a mSUGRA point, which gives 
the same mass of neutralino LSP as the benchmark point, in Fig.\ref{fig:masses}(b).
One can notice that the spectrum of the benchmark point is quite degenerated, 
compared to the mSUGRA point.

With these mass parameters for the benchmark point,
the neutralino LSP turns out to be bino-like, 
$but$ with non-negligible wino and Higgsino components.
Neutralino pair annihilation into a gauge boson pair 
$(ZZ~ {\rm or}~ W^+ W^-)$ is then efficient so that
thermal relic density of the neutralino LSP satisfies WMAP bound 
though the LSP is rather heavy ($m_{\tilde\chi^0_1} \sim 355$ GeV).

For the benchmark point, the ratio of gaugino masses at EW scale is given by 
$M_1 : M_2: M_3 = 367 : 461 : 850 \simeq 1 : 1.26 : 2.32$, 
which is quite degenerated comparing to the typical ratio 
$M_1 : M_2: M_3 \simeq 1: 2 : 6$ for mSUGRA-like scenarios in which gaugino masses are unified at
the GUT scale. For the benchmark point, 
the mass ratio of the gluino to the lightest neutralino is given by 
$m_{\tilde g}/m_{\tilde\chi_1^0} \simeq 2.5$, which is different from the typical 
mSUGRA predictions $m_{\tilde g}/m_{\tilde\chi_1^0} \gtrsim 6$ and therefore 
implies that the gaugino masses are NOT unified at the GUT scale. 
In the next subsection, we will see 
the mass ratio $m_{\tilde g}/m_{\tilde\chi_1^0}$ can be measured experimentally so that
we can get some information on SUSY breaking sector.

\subsection{Monte Carlo Events}

A Monte Carlo event sample of the SUSY signals for proton-proton collision at 
an energy of 14 TeV has been generated by PYTHIA 6.4 \cite{pythia}. 
The event sample corresponds to 30 $fb^{-1}$ of integrated luminosity,
which is expected with the 3 year running of the LHC at low luminosity.
We have also generated SM background events $i.e.,$ $t\bar t$ events equivalent to 30 $fb^{-1}$ of
integrated luminosity and also $W/Z + jet$, $WW/WZ/ZZ$ and QCD events, with less equivalent luminosity,
in five logarithmic $p_T$ bins for 50 GeV $< p_T <$ 4000 GeV.
The generated events have been further processed with 
a modified version of the fast detector simulation program PGS (Pretty Good Simulation)\cite{PGS}, 
which approximate an ATLAS- or CMS-like detector with reasonable efficiencies and fake rates.

The total production cross section for SUSY events at the LHC 
is $\sim$6.1 pb for the benchmark point,
corresponding to $\sim 1.8 \times 10^5$ events with 30 $fb^{-1}$ luminosity.
Squarks and gluinos are expected to be copiously produced at the LHC, 
if $m_{\tilde q}$, $m_{\tilde g} \lesssim 1$ TeV. With the masses of gluino and squarks
in the Eq. (\ref{mass1}), the production cross sections 
for the $\tilde g \tilde g$, $\tilde g \tilde q$,
and $\tilde q \tilde q$ pairs are about $0.3$ pb, $2.7$ pb and $2.0$ pb, respectively,
such that the gluino-squark pair production and the squark pair production dominate.

The produced squarks and gluinos decay generally in multistep. 
In the recent past, the following cascade decay chain of squark has been exploited 
in detail \cite{Bachacou:1999zb,weiglein};
\begin{eqnarray}
\tilde q_L \rightarrow \tilde\chi^0_2 q \rightarrow \tilde l^\pm_R~ l^\mp q
\rightarrow \tilde\chi^0_1 l^+ l^- q,
\label{squarkdecay}
\end{eqnarray}
especially in mSUGRA framework. It was shown that it would be possible to 
reconstruct both upper edges for the $l^+ l^-$, $l^+ l^- q$, and $l^\pm q$ mass distribution 
and a lower edge for the $l^+ l^- q$ mass coming from 
backwards decays of the $\tilde\chi^0_2$ in the $\tilde q_L$ rest frame. 
Those edge values of the mass distributions are given by 
the following analytic formulae, 
in terms of the particle masses involved in the decay chain; 
\begin{eqnarray}
M_{ll}^{max} = \left[\frac{(m_{\tilde\chi_2^0}^2 - m_{\tilde l_R}^2)
(m_{\tilde l_R}^2-m_{\tilde\chi_1^0}^2)}{m_{\tilde l_R}^2} \right]^{1/2},
\label{llmax}
\end{eqnarray}
\begin{eqnarray}
M_{llq}^{max} = \left[\frac{(m_{\tilde q_L}^2 - m_{\tilde\chi_2^0}^2)
(m_{\tilde\chi_2^0}^2-m_{\tilde\chi_1^0}^2)}{m_{\tilde\chi_2^0}^2} \right]^{1/2},
\end{eqnarray}
\begin{eqnarray}
M_{lq}^{max} (high) = \left[\frac{(m_{\tilde q_L}^2 - m_{\tilde\chi_2^0}^2)
(m_{\tilde\chi_2^0}^2-m_{\tilde l_R}^2)}{m_{\tilde\chi_2^0}^2} \right]^{1/2},
\end{eqnarray}
\begin{eqnarray}
(M_{llq}^{min})^2 &=&\frac{1}{4 m_{\tilde\chi_2^0}^2 m_{\tilde l_R}^2} 
\big[-m_{\tilde\chi_1^0}^2 m_{\tilde\chi_2^0}^4 + 3 m_{\tilde\chi_1^0}^2 
m_{\tilde\chi_2^0}^2 m_{\tilde l_R}^2
-m_{\tilde\chi_2^0}^4 m_{\tilde l_R}^2 -m_{\tilde\chi_2^0}^2 m_{\tilde l_R}^4 \nonumber\\
&-& m_{\tilde\chi_1^0}^2 m_{\tilde\chi_2^0}^2 m_{\tilde q_L}^2
- m_{\tilde\chi_1^0}^2 m_{\tilde l_R}^2 m_{\tilde q_L}^2
+3 m_{\tilde\chi_2^0}^2 m_{\tilde l_R}^2 m_{\tilde q_L}^2 - m_{\tilde l_R}^4 m_{\tilde q_L}^2 \\
&+&(m_{\tilde\chi_2^0}^2-m_{\tilde q_L}^2) 
\sqrt{(m_{\tilde\chi_1^0}^4+m_{\tilde l_R}^4)(m_{\tilde\chi_2^0}^2+m_{\tilde l_R}^2)^2
+2 m_{\tilde\chi_1^0}^2 m_{\tilde l_R}^2 (m_{\tilde\chi_2^0}^4-6 m_{\tilde\chi_2^0}^2 m_{\tilde l_R}^2
+m_{\tilde l_R}^4)}~ \big]. \nonumber
\end{eqnarray}
We can also measure the upper edge of the distribution of 
the $smaller$ of the two possible $lq$ masses formed by 
combining two leptons with a quark jet \cite{dkn00}; 
\begin{eqnarray}
M_{lq}^{max} (low)&=&\left[\frac{(m_{\tilde q_L}^2-m_{\tilde\chi_2^0}^2) (m_{\tilde e_R}^2-m_{\tilde\chi_1^0}^2)}
{2 m_{\tilde e_R}^2-m_{\tilde\chi_1^0}^2}\right]^{1/2}
~{\rm for}~ 2 m_{\tilde e_R}^2 - (m_{\tilde\chi_1^0}^2+m_{\tilde\chi_2^0}^2) < 0. \nonumber \\
&=&\left[\frac{(m_{\tilde q_L}^2-m_{\tilde\chi_2^0}^2) (m_{\tilde\chi_2^0}^2-m_{\tilde e_R}^2)}
{m_{\tilde\chi_2^0}^2}\right]^{1/2}
~{\rm for}~ 2 m_{\tilde e_R}^2 - (m_{\tilde\chi_1^0}^2+m_{\tilde\chi_2^0}^2) > 0. 
\label{lqmaxlow}
\end{eqnarray}
From the kinematic edge measurements, the SUSY particle masses might be then 
determined without relying on a model.
For the benchmark point, we can notice that 
$m_{\tilde\chi^0_2} > m_{\tilde e_R}$. 
Therefore, the cascade decay chain (\ref{squarkdecay}) is indeed open 
so that the well-established method of the kinematic edge measurements can be applied to our case.

In order to reduce the SM background to a negligible level, 
we apply the following event selection cuts;

(1) At least three jets with $P_{T1} > 200$ GeV and $P_{T2,3} > 50$ GeV.

(2) Missing transverse energy, $E_T^{miss} > 200~ {\rm GeV}.$

(3) $E_T^{miss}/M_{eff} > 0.2$, where 
$M_{eff} \equiv P_{T1}+P_{T2}+P_{T3}+P_{T4} + E_T^{miss} + \sum_{leptons} P_{Tl}$

(4) At least two isolated leptons of opposite charge, with $P_T > 10$ GeV and $|\eta|<2.5$. 

(5) Transverse sphericity $S_T > 0.1$.

(6) No b-jets.

\begin{figure}[ht!]
\vskip 0.6cm
\begin{center}
\epsfig{figure=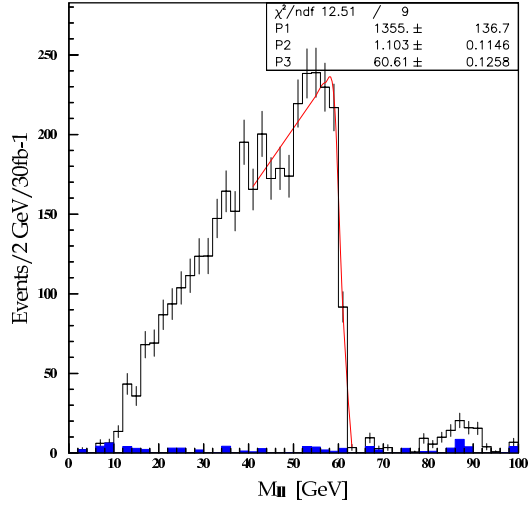,width=8cm,height=8cm,angle=270}
\end{center}
\vskip 0.1cm
\caption{\it Dilepton ($e^+e^-+\mu^+\mu^--e^\pm \mu^\mp$) 
invariant mass distributions for the benchmark point. Blue histogram corresponds to
SM background.}
\label{fig:dileptons}
\end{figure}

Following the analysis in Ref.\cite{Bachacou:1999zb},
we have then calculated various invariant mass distributions for the benchmark point. 
The dilepton invariant mass distribution is shown in the Fig. \ref{fig:dileptons}.
Here, the $e^+e^- + \mu^+\mu^- - e^\pm \mu^\mp$ combination was used in order to
cancel contributions for two independent decays and reduce combinatorial background.
The SM backgrounds which denoted as blue histogram on the plot, are negligible
with the above event selection cuts, as expected from previous studies 
\cite{LHC,Bachacou:1999zb}.
We find a clear end point in the dilepton mass distribution.
A Gaussian-smeared triangular fit to the distribution gives 
$M_{ll}^{max} = 60.61 \pm 0.13 $ GeV,
which is $\sim 0.3\%$ lower than the calculated value of $M_{ll}^{max}=60.8$ GeV
for the decays $\tilde\chi_2^0 \rightarrow \tilde l_R l^\pm \rightarrow 
\tilde\chi_1^0 l^+ l^-$.

\begin{figure}[ht!]
\vskip 0.6cm
\begin{center}
\epsfig{figure=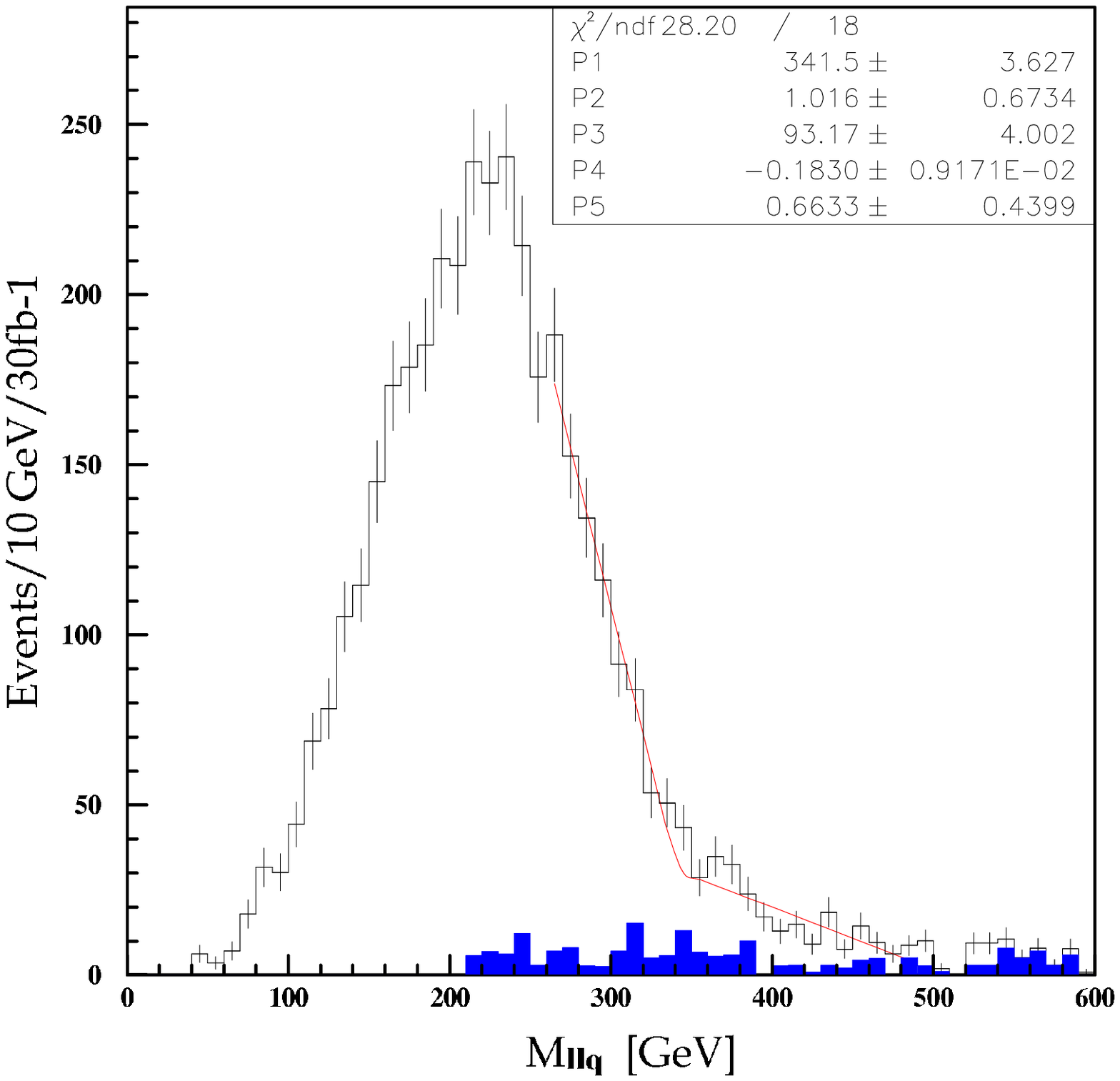,width=6cm,height=6cm,angle=270}
\epsfig{figure=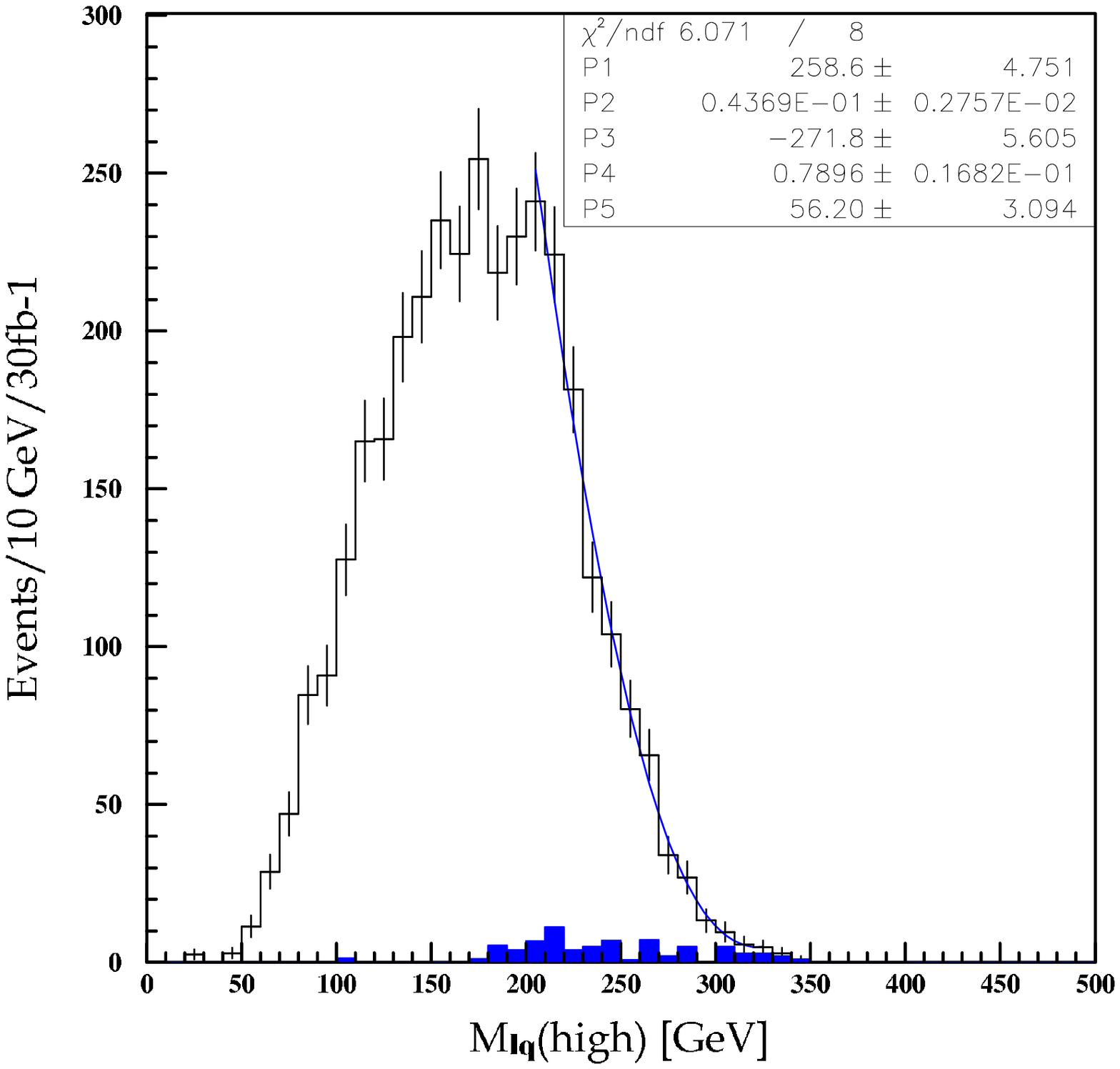,width=6cm,height=6cm,angle=270}
\epsfig{figure=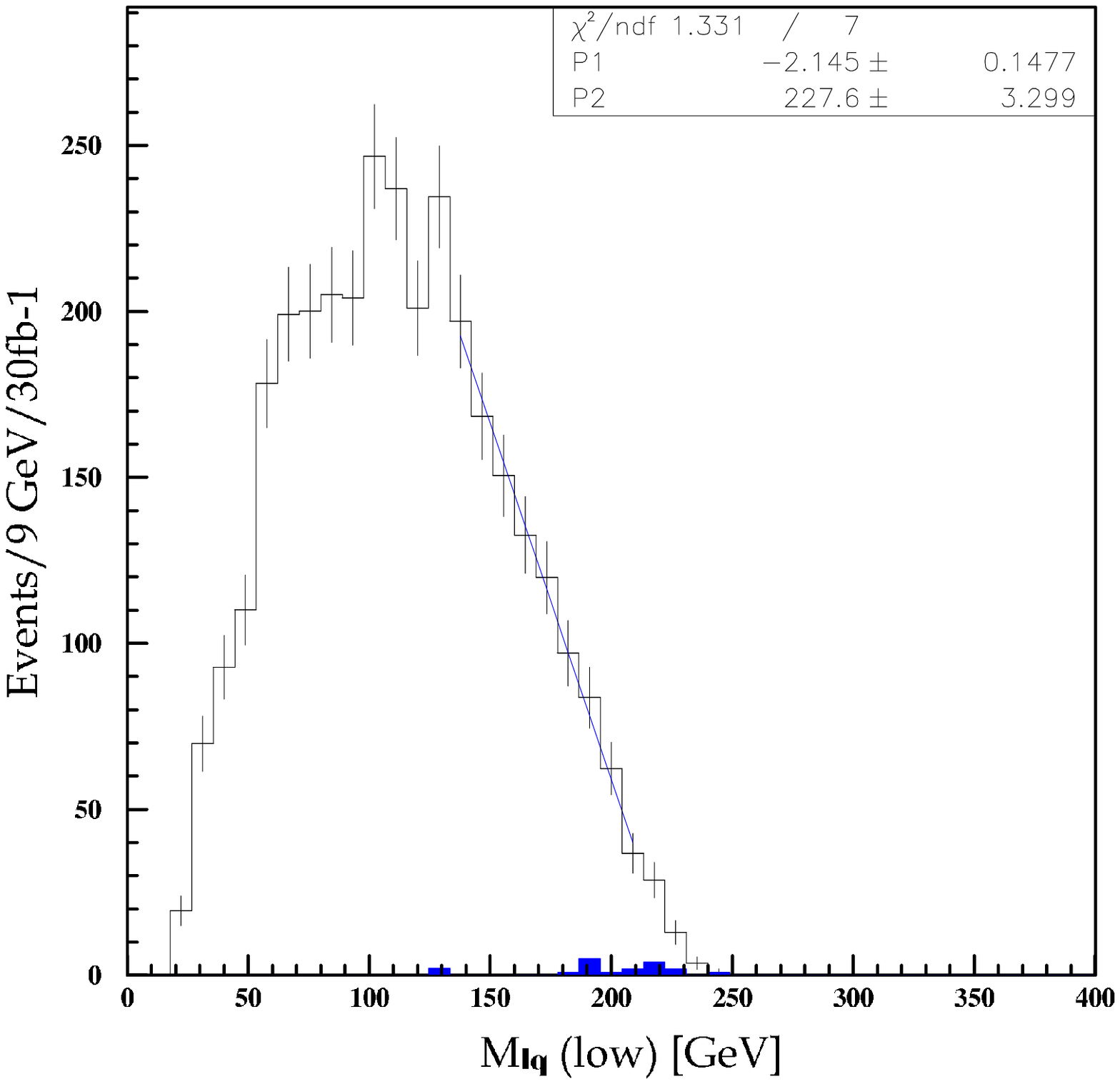,width=6cm,height=6cm,angle=270}
\epsfig{figure=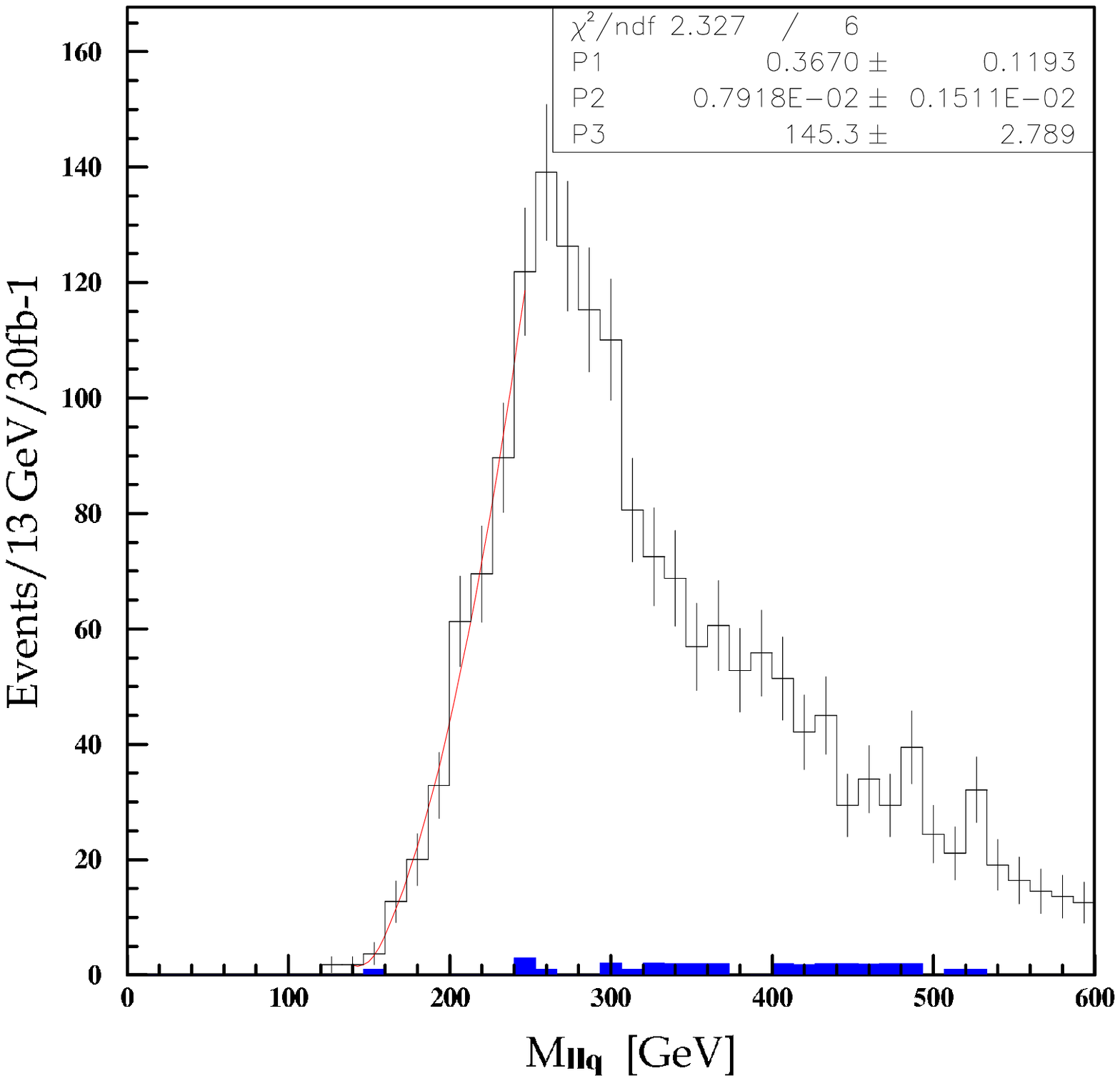,width=6cm,height=6cm,angle=270}
\end{center}
\vskip 0.4cm
\caption{\it Distributions for
(a) the smaller of the two $M_{llq}$, (b) the larger of two $M_{lq}$,
(c) the smaller of the two $M_{lq}$, and (d) the larger of the two $M_{llq}$
with $M_{ll}>M_{ll}^{max}/\sqrt{2}$.}
\label{fig:invmasses}
\end{figure}

The dilepton plus jet invariant mass $M_{llq}$ has been also calculated 
by combining the dilepton with one of the two hardest $P_T$ jets. 
The two hardest jets are expected to come from squark decays 
$\tilde q\rightarrow \tilde\chi q$ 
as dominant production processes result in a pair of squark .
The distribution for the smaller of the two possible $llq$ masses 
is shown in Fig. \ref{fig:invmasses}(a). 
The upper edge value of the $l^+l^-q$ masses is given by $M_{llq}^{max}=341.5 \pm 3.6$ GeV, 
which is obtained from a Gaussian smeared fit plus a linear background.
The fitted value is consistent with the calculated value of 
$M_{llq}^{max}=341.4$ GeV for the decays 
$\tilde q_L \rightarrow \tilde\chi_2^0 q \rightarrow \tilde l_R^\pm l^\mp q
\rightarrow \tilde\chi_1^0 l^+ l^- q$ (with $m_{\tilde q_L}=776$ GeV).

Now, we further require that one $M_{llq}$ should be less than 350 GeV and the other greater.
Two $l^\pm q$ masses are then calculated using the combination of $l^+l^-q$ 
with the smaller $M_{llq}~ (< 350~ {\rm GeV})$.
Fig. \ref{fig:invmasses}(b) shows the mass distribution of 
the $larger$ of the two $l^\pm q$ system.
A fit to the mass distribution near the end point is also shown, which yield 
an upper edge value $M_{lq}^{max}(high)=258.6 \pm 4.8$ GeV, which is consistent with 
the calculated value of $M_{lq}^{max}(high)=259.7$ GeV.

The mass distribution of the $smaller$ of the two $M_{lq}$ 
is shown in Fig. \ref{fig:invmasses}(c).
The upper edge value of the distribution from a linear fit is 
given by $M_{lq}^{max} (low) = 227.6 \pm 3.3$ GeV, which is consistent with
the calculated value of $M_{lq}^{max} (low) = 226.5$ GeV.

The lower edge for $l^+l^-q$ mass can be reconstructed 
from the larger of the two possible $l^+l^-q$ masses,
with additional requirement $M_{ll}^{max}/\sqrt{2} < M_{ll} < M_{ll}^{max}$. 
The resulting distribution is shown in Fig. \ref{fig:invmasses}(d).
A fit to the distribution gives $M_{llq}^{min}=145.3 \pm 2.8$ GeV, which is consistent with
the generated value $M_{llq}^{min}=145.4$ GeV.

From the above five edge measurements for the $\tilde q_L$ cascade decays (\ref{squarkdecay}), 
we can determine the masses of $\tilde q_L$, $\tilde l_R$, $\tilde\chi_2^0$ and
$\tilde\chi_1^0$ using the eqs. (\ref{llmax}-\ref{lqmaxlow}).

\begin{figure}[ht!]
\vskip 0.6cm
\begin{center}
\epsfig{figure=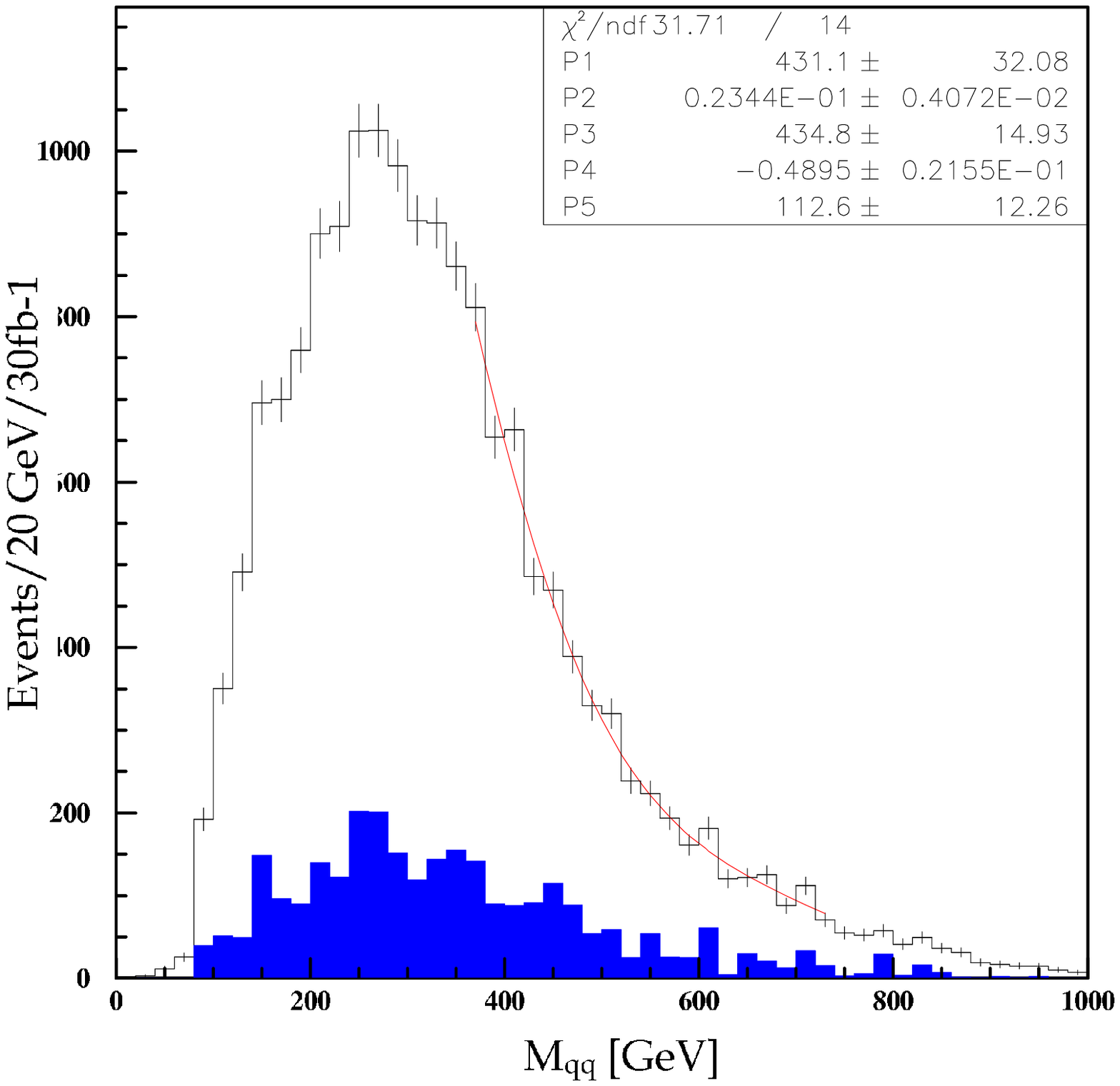,width=7cm,height=7cm,angle=270}
\epsfig{figure=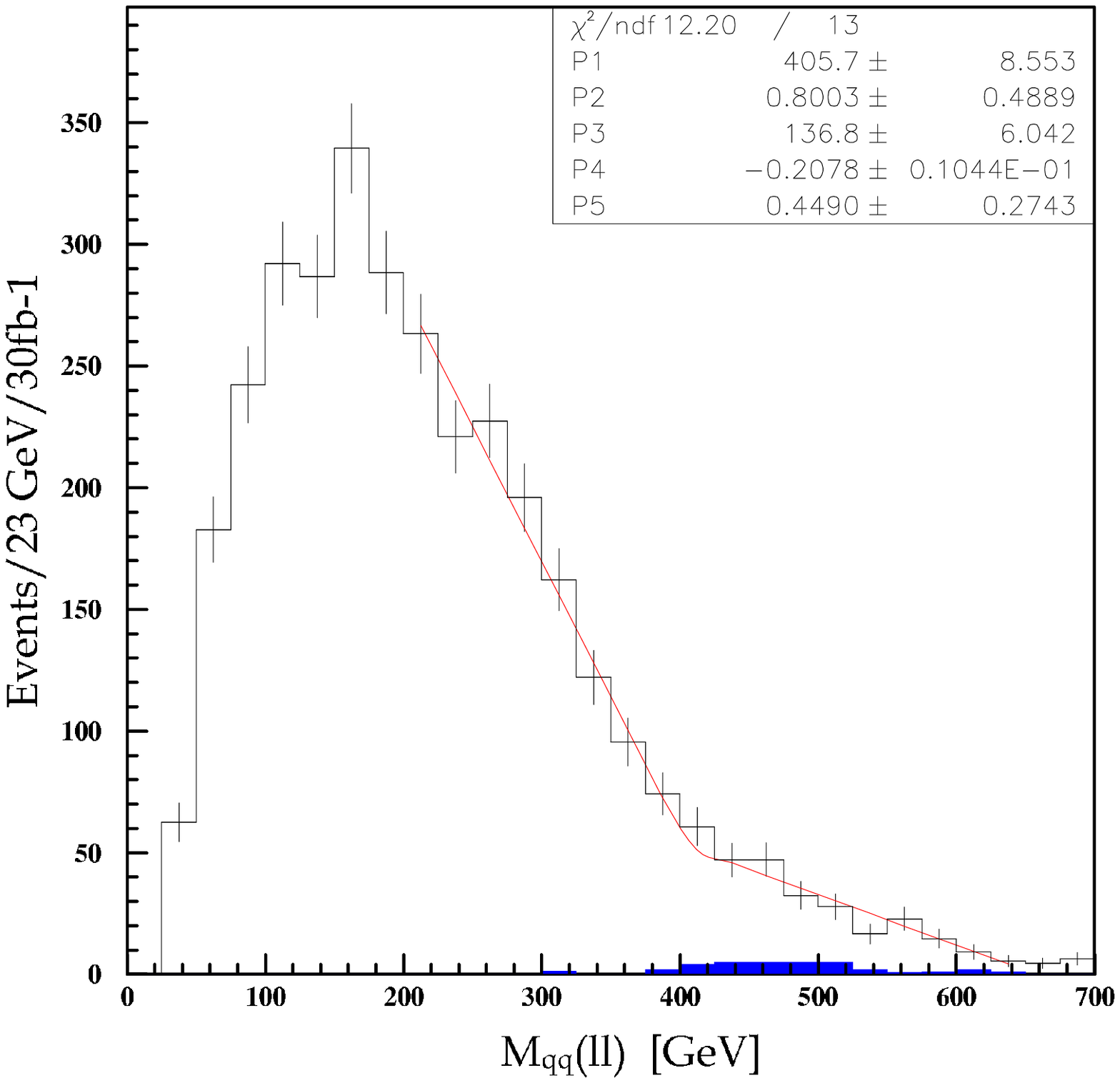,width=7cm,height=7cm,angle=270}
\end{center}
\vskip 0.4cm
\caption{\it Dijet invariant mass distribution (a) without dilepton and (b) with dilepton.}
\label{fig:dijet}
\end{figure}

With given masses of $\tilde\chi_1^0$ and $\tilde\chi_2^0$, 
we can get information for gluino 
and squark masses from the following gluino decay into right-handed squark plus quark;
\begin{eqnarray}
\tilde g \rightarrow \tilde q_R q \rightarrow \tilde\chi_1^0 q q.
\label{gluino1}
\end{eqnarray}
The maximal value of the two jet invariant mass for the gluino decays (\ref{gluino1}) is given by
\begin{eqnarray}
M_{qq}^{max}=
\left[\frac{(m_{\tilde g}^2 - m_{\tilde q_R}^2)
(m_{\tilde q_R}^2-m_{\tilde\chi_1^0}^2)}{m_{\tilde q_R}^2} \right]^{1/2}. 
\end{eqnarray}
At the LHC, gluino would be produced mainly from 
$\tilde g \tilde q$ or $\tilde g \tilde g$ pair production.
In order to obtain event sample for the gluino decay (\ref{gluino1}), 
the following event selection cuts are imposed;

(1) At least 3 jets with $P_{T1}> 200$ GeV, $P_{T2} > 150$ GeV and $P_{T3} > 100$ GeV.

(2) $E_T^{miss} > 350$ GeV and $E_T^{miss}/M_{eff} > 0.25$.

(3) Transverse sphericity $S_T > 0.15$.

(4) No b-jets and No leptons.

In the selection cuts, we required rather hard $P_{T3} > 100$ GeV in order to reduce 
${\tilde q}{\tilde q}$ production events.
After the selection cuts, we calculate dijet invariant mass $M_{qq}$ using the three hardest jets.
The smallest $M_{qq}$ among the three possible combinations 
is then shown in Fig. \ref{fig:dijet} (a).
The $M_{qq}$ distribution is fitted near end point by a Gaussian smeared linear function 
with a linear background. The resulting edge value is given by 
$M_{qq}^{max}=431.1 \pm 32.1$ GeV,
which is consistent with 
the calculated one $M_{qq}^{max} = 432.4$ GeV (with $m_{\tilde q_R}=733.5$ GeV).


Additional information on $m_{\tilde g}$ and $m_{\tilde q_R}$ might be provided
by the following cascade decay of gluino,
\begin{eqnarray}
\tilde g \rightarrow \tilde q_R q \rightarrow \tilde\chi_2^0 q q 
~(\rightarrow \tilde\chi_1^0 l l q q),
\label{gluino2}
\end{eqnarray}
for which the upper edge value of the two jet invariant mass distribution is given by
\begin{eqnarray}
M_{qq}^{max} (ll)=
\left[\frac{(m_{\tilde g}^2 - m_{\tilde q_R}^2)
(m_{\tilde q_R}^2-m_{\tilde\chi_2^0}^2)}{m_{\tilde q_R}^2} \right]^{1/2}. 
\label{mqq2}
\end{eqnarray}
In the gluino cascade decay (\ref{gluino2}), we consider the right-handed squark decay into 
$\tilde\chi_2^0$ rather than $\tilde\chi_1^0$, where
the $\tilde\chi_2^0$ further undergoes the dileptonic decay.
In the benchmark point, branching ratio
$BR(\tilde q_R \rightarrow \tilde\chi_2^0 q)=11\%$ is comparable to 
$BR(\tilde q_L \rightarrow \tilde\chi_2^0 q)=18.5\%$.
This is because the gaugino masses are quite degenerated 
$(i.e.~M_2/M_1 \sim 1.26)$ at EW scale
so that $\tilde\chi_2^0$ has sizable bino-component 
in the benchmark point, which is in contrast to the typical mSUGRA case where
$\tilde\chi_2^0$ is almost wino-like and therefore
$BR(\tilde q_R \rightarrow \tilde\chi_2^0 q)$ is negligible. 
Furthermore, the gluino branching ratio of $BR(\tilde g \rightarrow \tilde q_R q)\simeq 19\%$
is larger than $BR(\tilde g \rightarrow \tilde q_L q)\simeq 11.2\%$ for q=u,d so that 
the decay chain of $\tilde g \rightarrow \tilde q_R q \rightarrow \tilde\chi_2^0 q q$ 
is comparable to that of $\tilde g \rightarrow \tilde q_L q \rightarrow \tilde\chi_2^0 q q$.
Considering the squark masses that $m_{\tilde q_R} < m_{\tilde q_L}$,
the upper edge value of two jet invariant mass for 
$\tilde\chi_2^0 q q$ events is essentially determined by the Eq. (\ref{mqq2}), which
involves $m_{\tilde q_R}$ rather than $m_{\tilde q_L}$.

In order to select events which include decay chain (\ref{gluino2}), we require

(1) At least 3 jets with $P_{T1} > 100$ GeV and $P_{T2,3}>50$ GeV

(2) $E_T^{miss} > 200$ GeV and $M_{eff}/E_T^{miss} > 0.2$

(3) At least two isolated leptons with opposite charge

(4) $M_{ll} < 61$ GeV and the smallest $M_{llq} < 350$ GeV

(5) Transverse sphericity $S_T > 0.1$

(6) No b-jets

After the selection cut, 
dijet invariant mass was calculated using the jet, which gives the smallest $M_{llq}$,
with third and fourth (if any) energetic jets. 
The smaller of the two possible dijet mass is then shown in Fig. \ref{fig:dijet} (b).
A fit to the distribution gives 
$M_{qq}^{max} (ll)=405.7 \pm 8.6$ GeV, which is consistent with
the calculated value of $M_{qq}^{max} (ll)= 406.9$ GeV (with $m_{\tilde q_R}=733.5$ GeV).

The upper edge measurements of two dijet invariant masses provide lower limits on
the gluino and squark masses 
and a strong correlation between $m_{\tilde g}$ and $m_{\tilde q_R}$,
but without constraint on the upper limit on the masses.
\begin{figure}[ht!]
\vskip 0.6cm
\begin{center}
\epsfig{figure=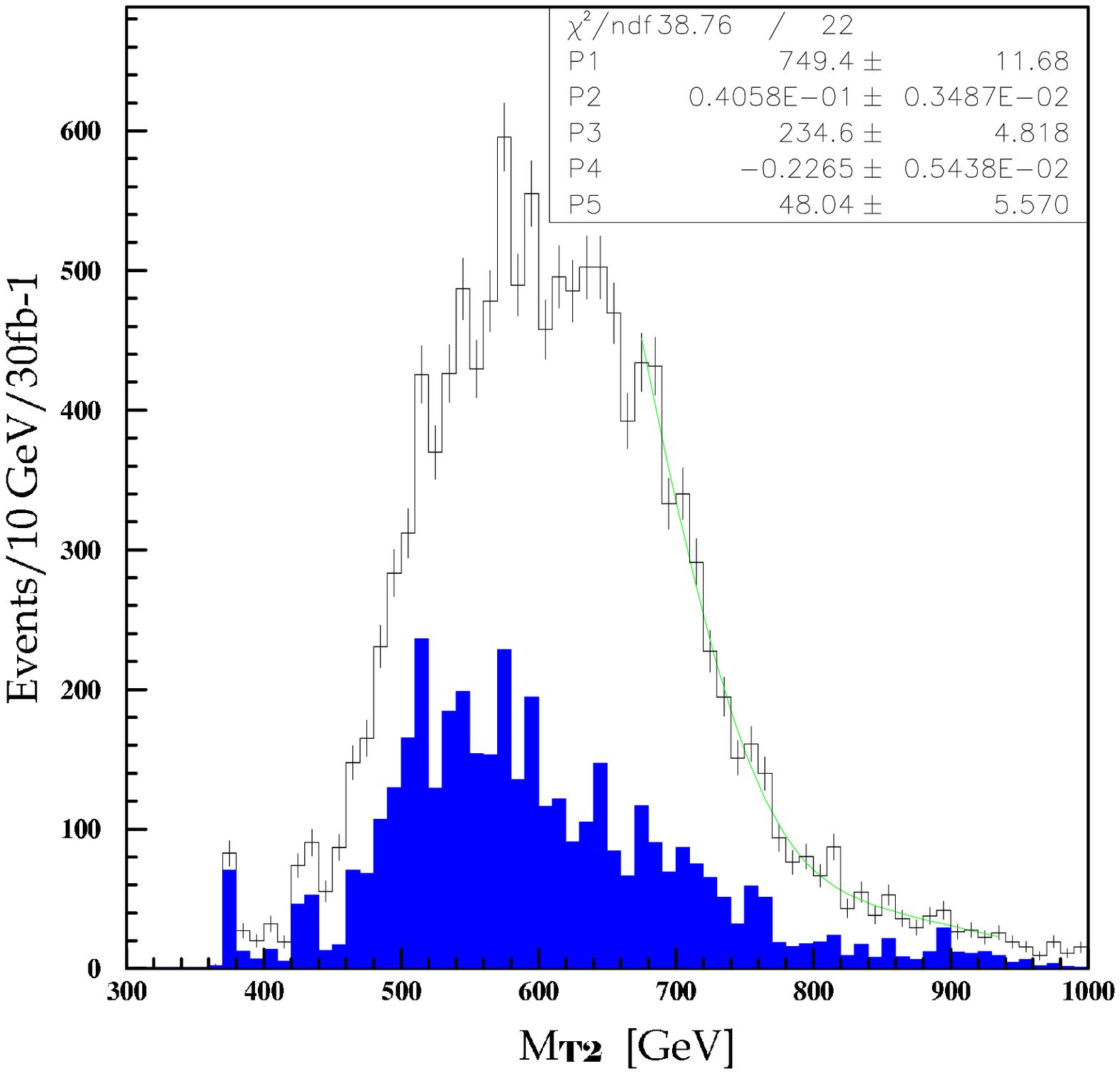,width=7cm,height=7cm,angle=270}
\epsfig{figure=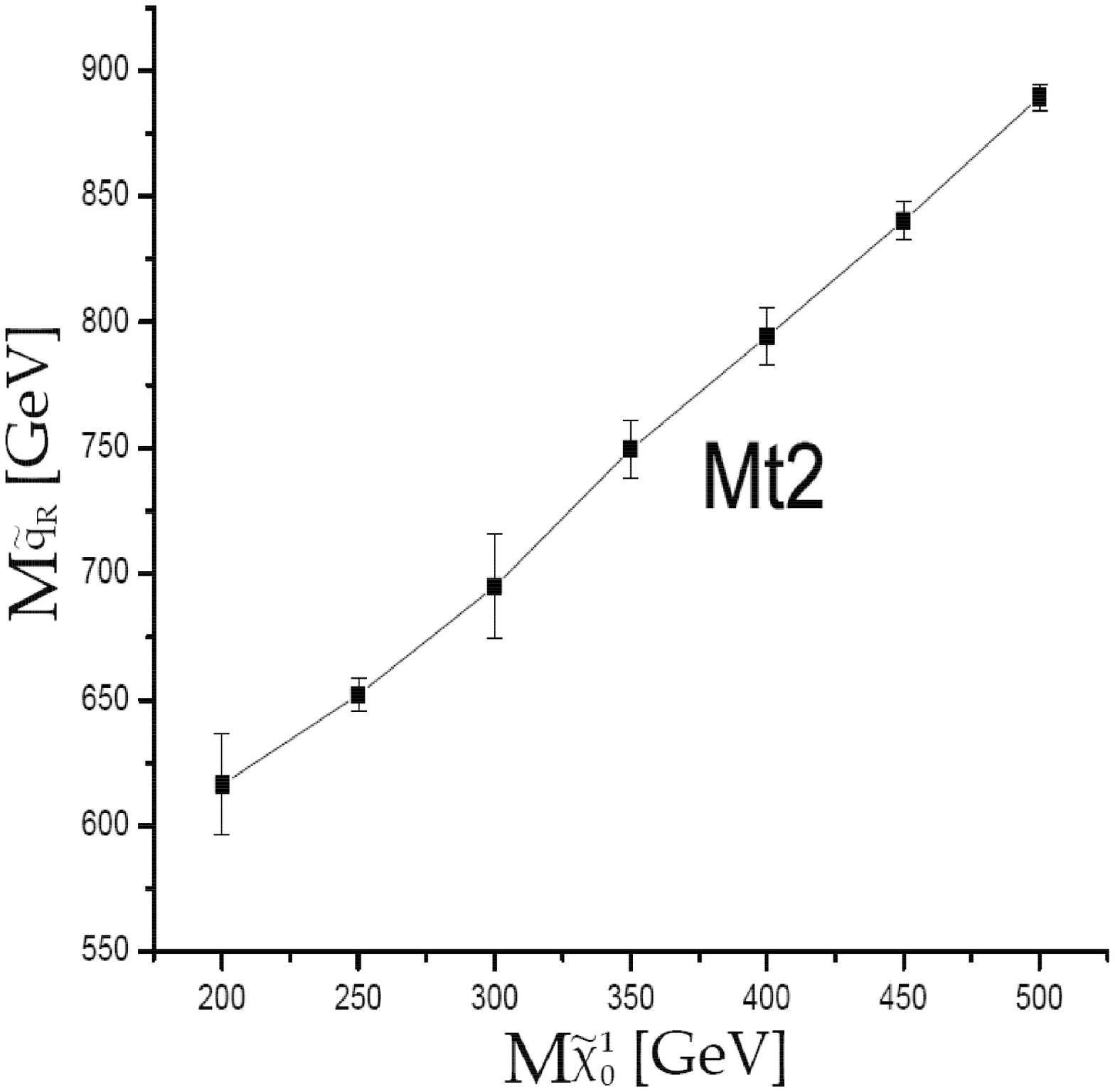,width=6.9cm,height=6.9cm,angle=270}
\end{center}
\vskip 0.4cm
\caption{\it (a) The $M_{T2}$ distribution with the input value of $m_{\chi}=350$ GeV
as an example. (b) A relation between $m_{\tilde q_R}$ and $m_{\tilde\chi_1^0}$ from 
the $M_{T2}$ analysis.}
\label{fig:mt2}
\end{figure}
In order to further constrain the right-handed squark mass, 
we consider squark pair production and their subsequent decay into 
two quarks plus two LSPs; 
\begin{eqnarray}
{\tilde q_{1R}}{\tilde q_{2R}} \rightarrow 
q_1 {\tilde\chi}_1^0 q_2 {\tilde\chi}_1^0,
\label{squarkpair}
\end{eqnarray}
and construct another variable $M_{T2}$ \cite{lester00},
which is defined by
\begin{eqnarray}
M_{T2}^2 \equiv \min_{{\bf p}_{T1}^\chi+{\bf p}_{T2}^\chi={\bf p}_T^{miss}} 
\left[ {\rm max} 
\{ m_T^2 ({\rm {\bf p}}_T^{q_1},{\rm {\bf p}}_{T1}^\chi), 
m_T^{2}({\rm {\bf p}}_T^{q_2},{\rm {\bf p}}_{T2}^\chi) \}  \right],
\end{eqnarray}
where 
\begin{eqnarray}
m_{T2}^2 ({\bf p}_T^q,{\bf k}_T^\chi) \equiv m_q^2 + m_{\chi}^2 + 
2 (E_T^q E_T^\chi - {\bf p}_T^q \cdot {\bf k}_T^\chi),
\end{eqnarray}
\begin{eqnarray}
E_T^q=\sqrt{|{\bf p}_T^q|^2 + m_q^2},~~E_T^\chi=\sqrt{|{\bf k}_T^\chi|^2 + m_\chi^2}.
\end{eqnarray}
Here, ${\bf p}_T^{q_1}$ and ${\bf p}_T^{q_2}$ denote 
the transverse momentum of quark-jets from the squark decays
and ${\bf p}_T^{miss}$ is the observed missing transverse momentum and $m_\chi$ is an estimate of
the lightest neutralino mass.
The $M_{T2}$ distribution has an endpoint at the squark mass $m_{\tilde q_R}$ 
when the input $m_{\chi}$ is equal to the correct $m_{\tilde\chi_1^0}$ value.
In general we obtain a relation between $m_{\tilde q_R}$ and $m_{\tilde\chi_1^0}$.

In order to have event sample for 
the squark pair production and their subsequent decay (\ref{squarkpair}), 
the following event selection cuts are required;

(1) At least two jets with $P_{T1} > 300$ GeV, $P_{T2} > 50$ GeV

(2) $E_T^{miss}>200$ GeV, $M_{eff}/E_T^{miss} > 0.3$

(3) Transverse sphericity $S_T > 0.15$

(4) No leptons, no b-jets.

Fig. \ref{fig:mt2}(a) shows an example of $M_{T2}$ distribution with an input value of $m_\chi=350$ GeV.
From a fit to the distribution, we obtain $m_{\tilde q_R} = 749 \pm 12$ GeV for the example.
A general relation between $m_{\tilde q_R}$ and $m_{\tilde\chi_1^0}$ from the $M_{T2}$ analysis is shown
in Fig. \ref{fig:mt2}(b).
\begin{figure}[ht!]
\vskip 0.0cm
\begin{center}
\epsfig{figure=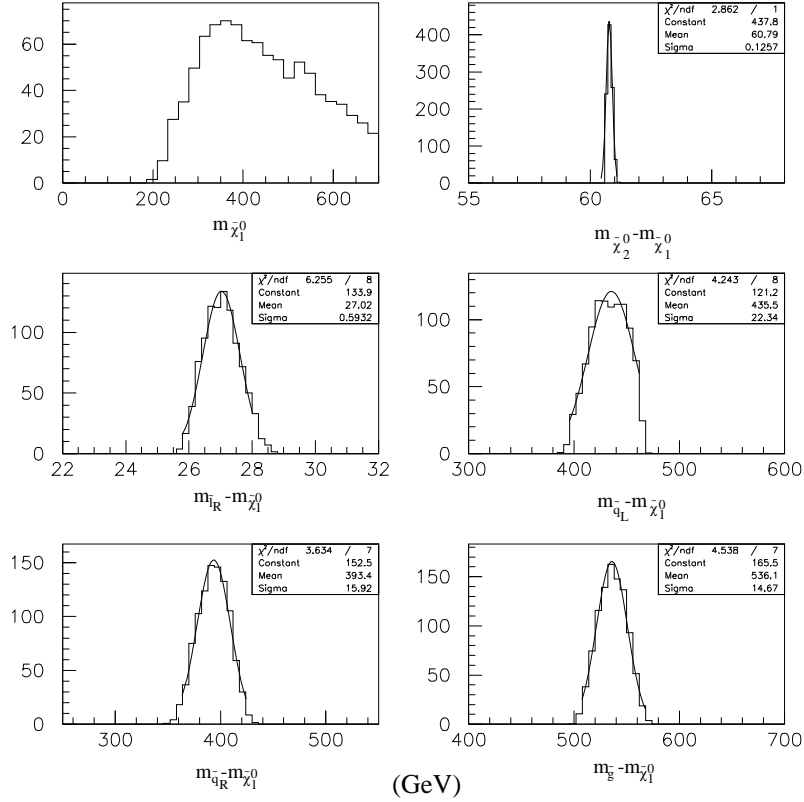,width=12cm,height=12cm}
\end{center}
\vskip 0.0cm
\caption{\it Probability distribution of (a) $m_{\tilde\chi_1^0}$, 
(b) $m_{\tilde\chi_2^0}-m_{\tilde\chi_1^0}$, (c) $m_{\tilde l_R}-m_{\tilde\chi_1^0}$,
(d) $m_{\tilde q_L}-m_{\tilde\chi_1^0}$, (e) $m_{\tilde q_R}-m_{\tilde\chi_1^0}$ and
(f) $m_{\tilde g}-m_{\tilde\chi_1^0}$.}
\label{fig:qlfit}
\end{figure}

Now we can determine six SUSY particle masses, $m_{\tilde\chi_1^0}$,
$m_{\tilde\chi_2^0}$, $m_{\tilde l_R}$, $m_{\tilde q_L}$, $m_{\tilde g}$ and $m_{\tilde q_R}$,
from the various kinematic distributions we considered so far.  
In order to scan possible values of SUSY particle masses,
random numbers for $m_{\tilde\chi_2^0}$, $m_{\tilde l_R}$, $m_{\tilde q_L}$, $m_{\tilde q_R}$,
and $m_{\tilde g}$ were generated within some ranges around their nominal values, 
while $m_{\tilde\chi_1^0}$ values were calculated with the measured $M_{ll}^{max}$ value.
The chi-square from the various kinematic observables with their errors 
was calculated to determine the probability for each set of masses. 
Fig.~\ref{fig:qlfit} shows the probability distribution of (a) $m_{\tilde\chi_1^0}$, 
(b) $m_{\tilde\chi_2^0}-m_{\tilde\chi_1^0}$, (c) $m_{\tilde l_R}-m_{\tilde\chi_1^0}$
(d) $m_{\tilde q_L}-m_{\tilde\chi_1^0}$, (e) $m_{\tilde q_R}-m_{\tilde\chi_1^0}$
and (f) $m_{\tilde g}-m_{\tilde\chi_1^0}$, respectively.
While the $\tilde\chi_1^0$ mass is determined as 
\begin{eqnarray}
m_{\tilde\chi_1^0} = 356_{-95}^{+220}~ {\rm GeV}, 
\end{eqnarray} 
the differences between sparticle masses are rather well
constrained, 
\begin{eqnarray}
m_{\tilde\chi_2^0}-m_{\tilde\chi_1^0}&=&60.8\pm0.1~ {\rm GeV}, 
~m_{\tilde l_R}-m_{\tilde\chi_1^0} =27.0 \pm 0.6~ {\rm GeV}, 
~m_{\tilde q_L}-m_{\tilde\chi_1^0}=436 \pm 22~ {\rm GeV}, \nonumber \\
~m_{\tilde q_R}-m_{\tilde\chi_1^0}&=&393 \pm 16~ {\rm GeV},
~m_{\tilde g}-m_{\tilde\chi_1^0}=536 \pm 15~ {\rm GeV},
\end{eqnarray} respectively. 
The central values for the estimated masses are consistent with the generated ones.
And the range for the mass ratio between gluino and the lightest neutralino 
is given by
\begin{eqnarray}
1.9 \lesssim \frac{m_{\tilde g}}{m_{\tilde\chi_1^0}} \lesssim 3.1  ~~~
(reduced~ \chi^2 <1),
\end{eqnarray}
which is quite distinctive from the typical predictions $i.e.,$
$m_{\tilde g}/m_{\tilde\chi_1^0} \gtrsim 6$
of the other SUSY scenarios in which gaugino masses are unified at GUT scale.

\begin{figure}[ht!]
\vskip 0.6cm
\begin{center}
\epsfig{figure=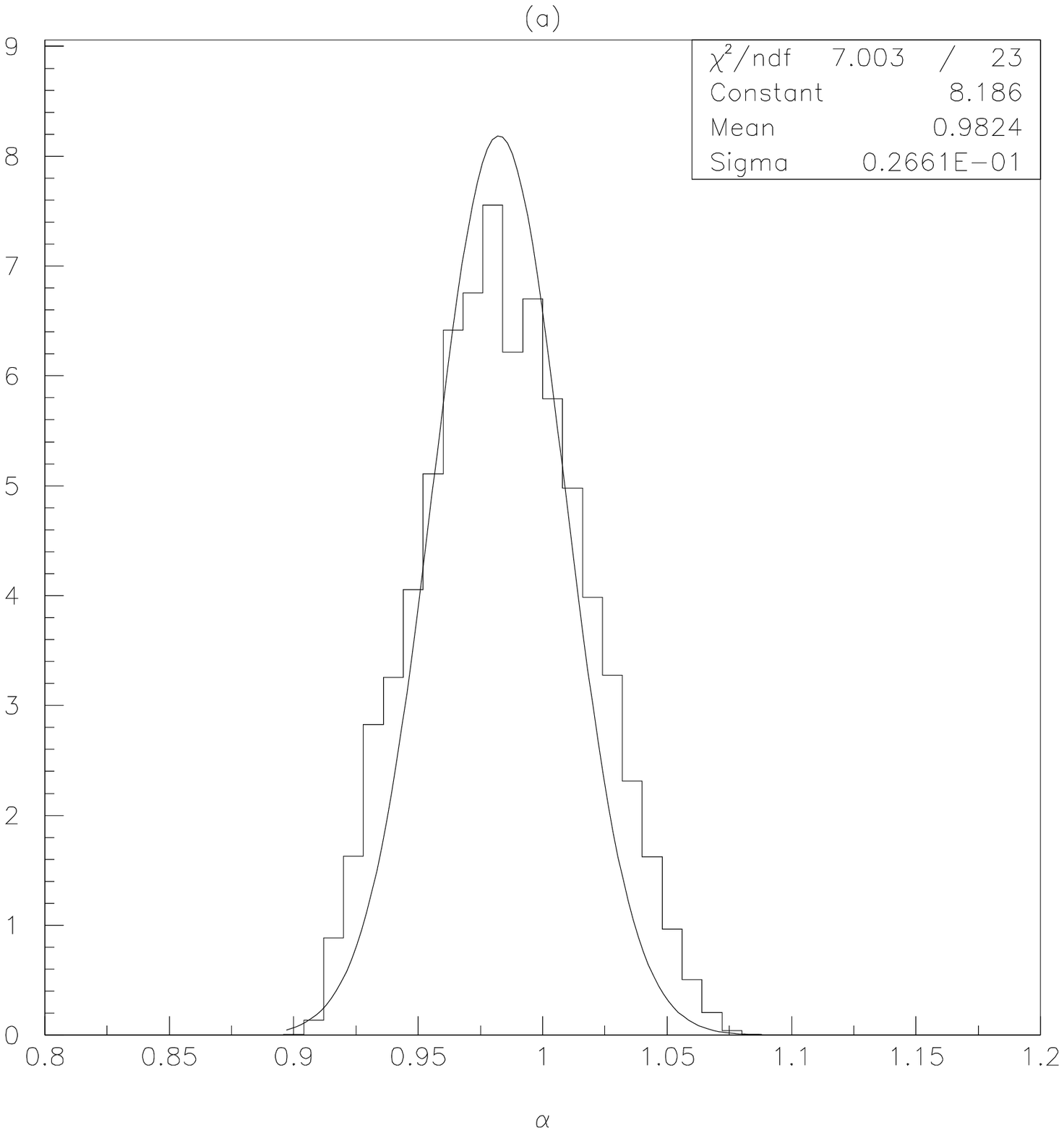,width=6cm,height=6cm}
\epsfig{figure=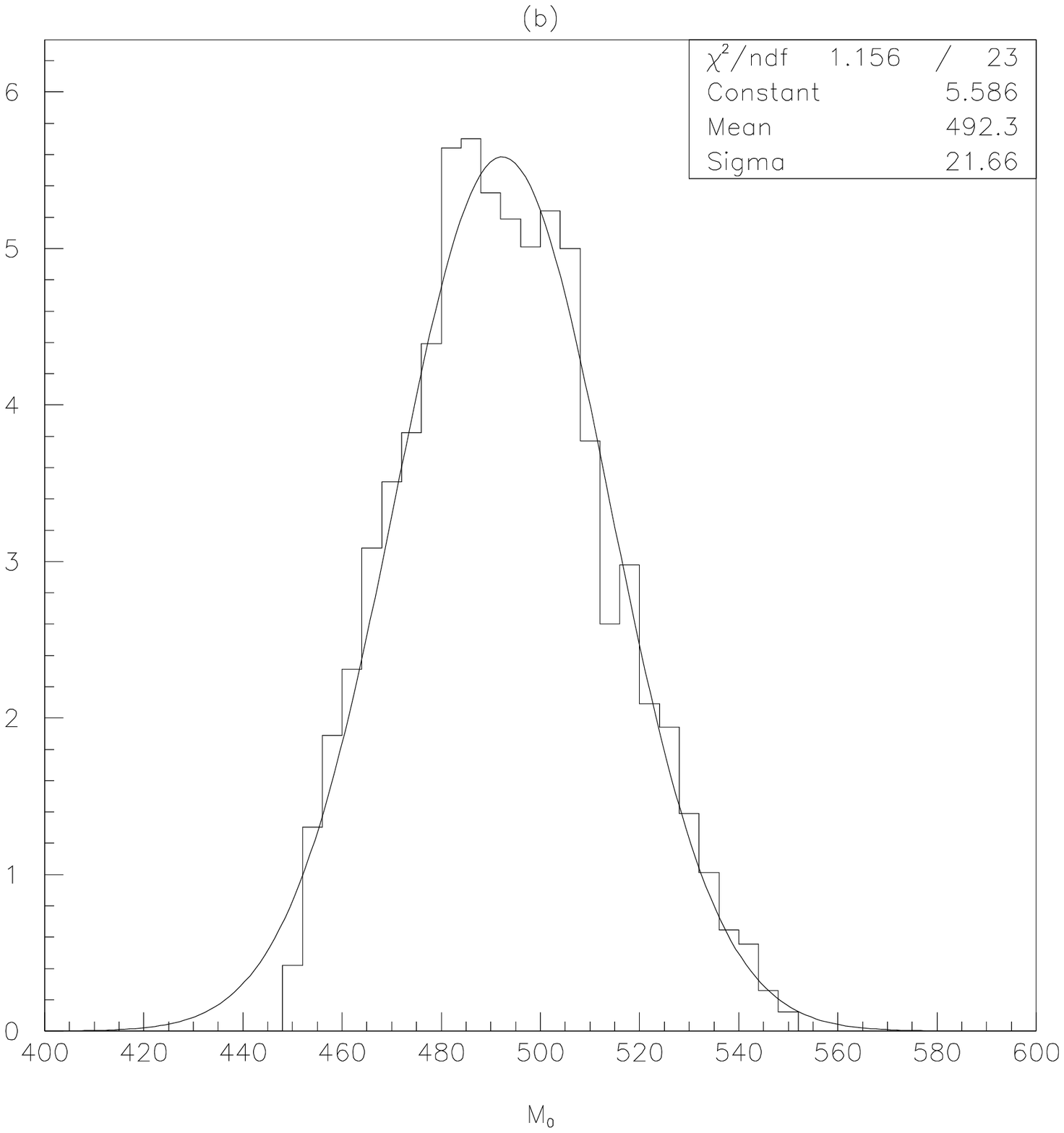,width=6cm,height=6cm}
\epsfig{figure=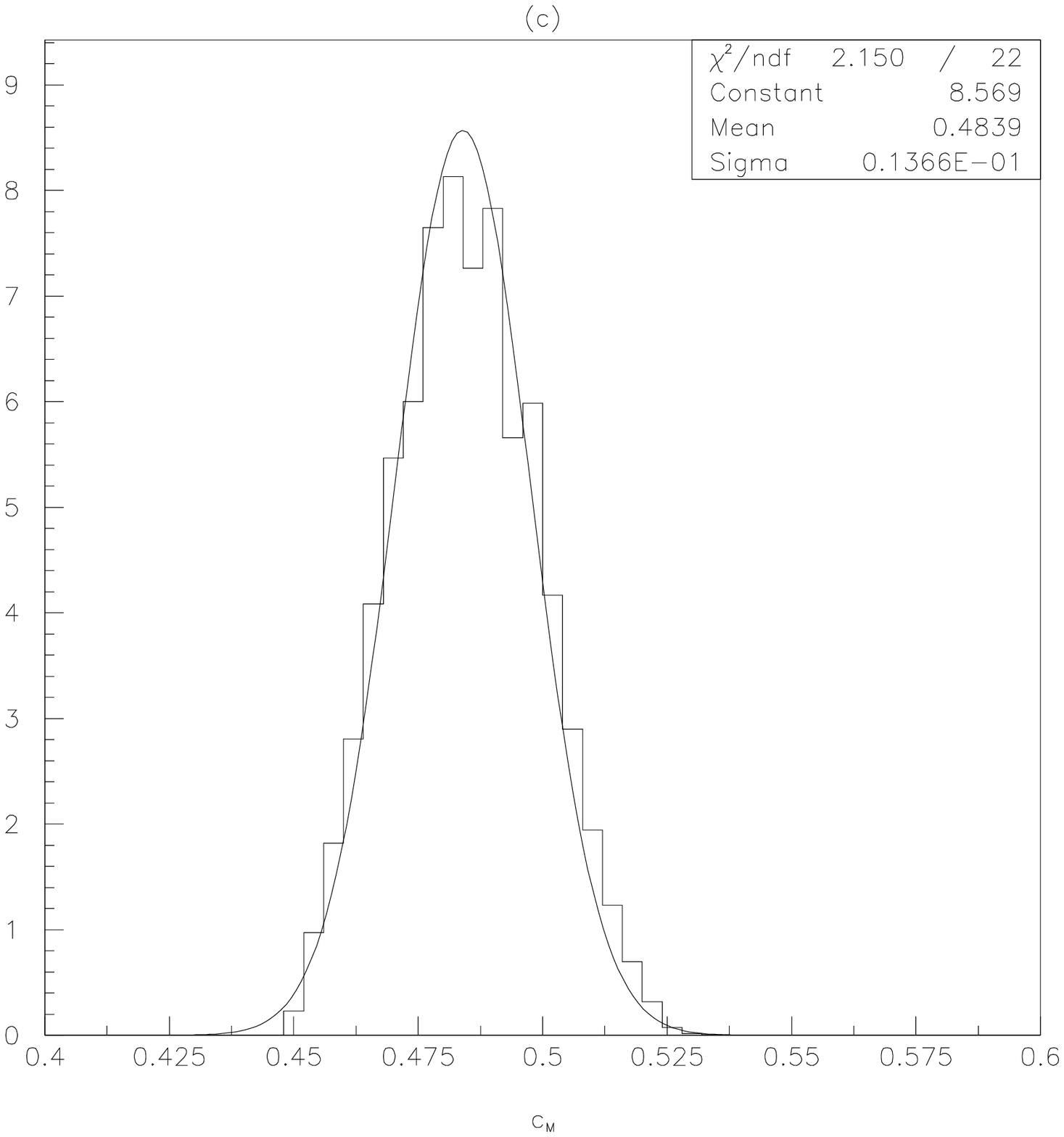,width=6cm,height=6cm}
\epsfig{figure=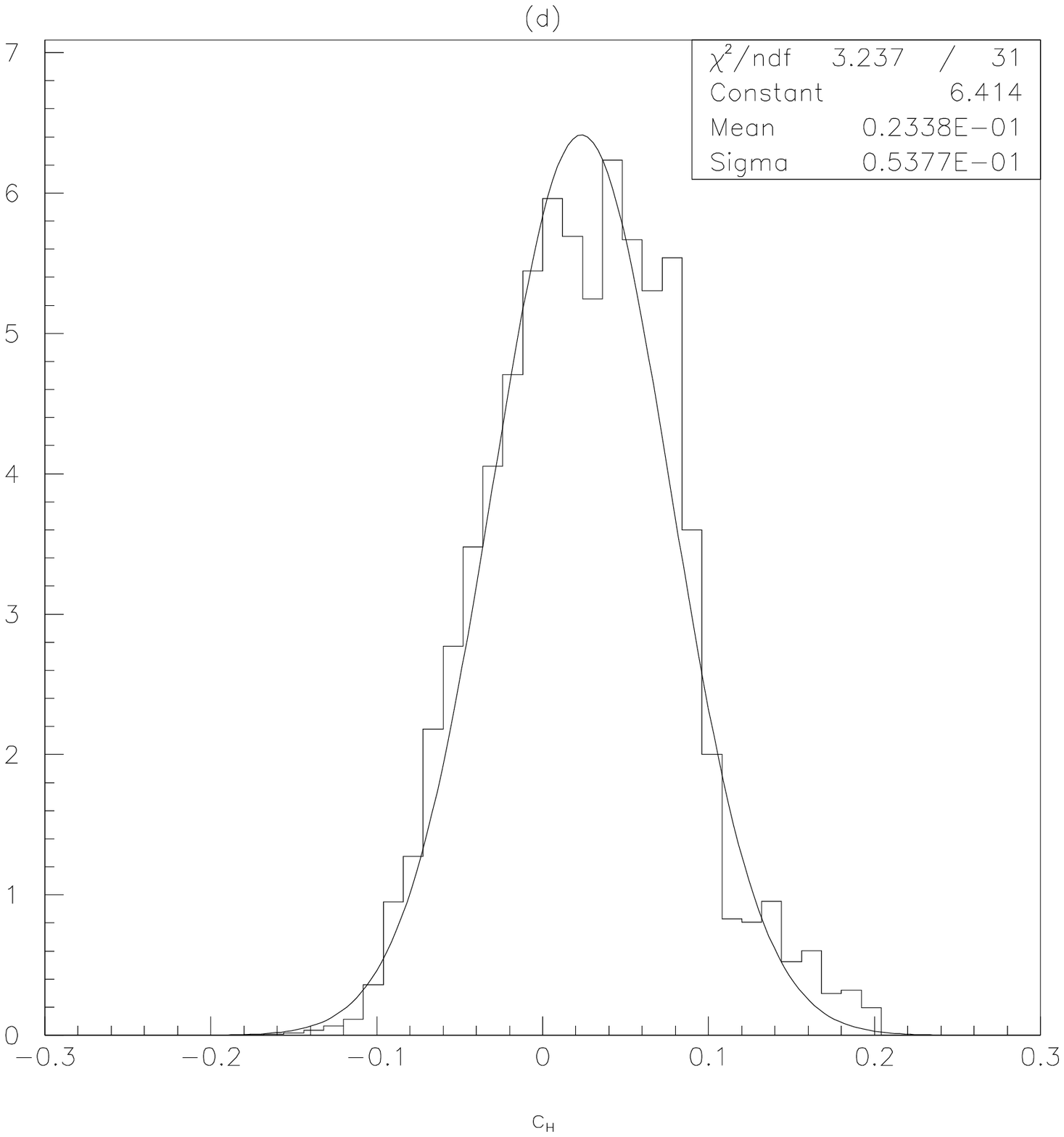,width=6cm,height=6cm}
\end{center}
\vskip 0.4cm
\caption{\it Probability distribution of (a) $\alpha$, (b) $M_0$, (c) $c_M$, and (d) $c_H$.}
\label{fig:modelfit}
\end{figure}

So far we have considered 'model-independent' measurements of 
SUSY particle masses.
We can also determine the model parameters of mirage 
mediation scenario from the kinematic endpoint measurements.
As mentioned in section 2, generic mirage mediation is parameterized by
$M_0, \alpha, a_i, c_i$ and ${\rm tan}\beta$. Here, we assume that visible sfermion fields
have a universal parameter $c_M(=a_M)$ 
while the corresponding parameter for Higgs fields is given by
$c_H(=a_H)$. The model is then described by 
five parameters $M_0, \alpha, c_M, c_H$ and ${\rm tan}\beta$.
From the Eq. (\ref{gauginomass}) and (\ref{scalarmass}) we can see that 
the gaugino and the first generation sfermion masses at low energy scale are essentially
determined by $M_0, \alpha$ and $c_M$. Therefore, we can determine the model parameters
$M_0, \alpha$ and $c_M$ from the measured $m_{\tilde g}, m_{\tilde q}$ and $m_{\tilde l_R}$.
With given values of $M_0, \alpha$ and $c_M$, 
the parameters $c_H$ and ${\rm tan}\beta$ might be obtained from 
the measured $m_{\tilde\chi_1^0}$ and $m_{\tilde\chi_2^0}$. 
We have scanned five model parameters in certain ranges and calculated the $\chi^2$ from
the calculated and measured kinematic edge values, in order to constraint the model parameters.
Fig. \ref{fig:modelfit} shows probability distributions for $\alpha$,$M_0$,$c_M$
and $c_H$, resulting in the following values of the parameters:
\begin{eqnarray}
\alpha = 0.98 \pm 0.03,~M_0= 492 \pm 22~ {\rm GeV},~ c_M = 0.48 \pm 0.01,
~c_H=0.02 \pm 0.05.
\end{eqnarray}
The model parameters are determined quite accurately, except ${\rm tan}\beta$ whose value
is not so well constrained because neutralino masses depend on it rather mildly.
The resulting central values of the parameters
well agree with the input values of Eq (\ref{modelparameters}).

\section{Conclusions}
\label{sec:conclusion}

In this paper we have investigated LHC signatures of mirage mediation by 
performing a Monte Carlo study for a benchmark point
in which anomaly and modulus contribution to 
soft SUSY breaking terms are comparable to each other.

The benchmark point has characteristic features of mirage mediation such as
rather degenerated gaugino masses at EW scale ($M_1 : M_2 : M_3 \simeq 1 : 1.26 : 2.3$)
and enhanced Wino- and Higgsino- components of the lightest neutralino, comparing to
typical mSUGRA case. Non-negligible wino and Higgsino components of the LSP leads to
right amount of thermal relic density which is compatible to WMAP data, though 
the LSP mass is rather large ($m_{\chi_1^0} \sim 355~ {\rm GeV}$). 
Rather degenerated gaugino masses at EW scale lead to the mass ratio of gluino to the lightest
neutralino $m_{\tilde g}/m_{\tilde\chi_1^0}\sim 2.5$, which is quite distinctive from
the typical predictions $i.e.,$ $m_{\tilde g}/m_{\tilde\chi_1^0} \gtrsim 6$ of the SUSY scenarios
which have gaugino mass unification at the GUT scale.

For the benchmark point, the `golden' decay chain of squark (\ref{squarkdecay}) is open, 
allowing model independent measurement of SUSY particle masses such as 
$m_{\tilde q_L}, m_{\tilde l_R}, 
m_{\tilde\chi_2^0}$ and $m_{\tilde\chi_1^0}$.
We also determined gluino and squark masses from various gluino and squark decays.
The SUSY particle masses are then determined in a model independent way, 
giving valuable information on SUSY breaking sector. In particular, the measured ratio
of $1.9 \lesssim m_{\tilde g}/m_{\tilde\chi_1^0} \lesssim 3.1$ 
well reproduce the theoretical input value $m_{\tilde g}/m_{\tilde\chi_1^0} \simeq 2.5$
of the benchmark point.
Therefore, the benchmark scenario may be distinguishable experimentally
from other SUSY scenario in which gaugino masses are unified at the GUT scale.
Model parameters were also obtained with small errors, 
from a global fit to the observables with their estimated errors. 
The resulting central values of the model parameters well agree with
the input values for the benchmark point.

\section*{Acknowledgements}

The authors thank Kiwoon Choi for useful discussions. 
This work was supported by 
the BK21 program of Ministry of Education (W.S.C., K.Y.L., C.B.P. and Y.S.), 
the Astrophysical Research Center for the Structure and Evolution 
of the Cosmos funded by the KOSEF (Y.G.K.),
the KRF Grant KRF-2005-210-C000006 funded by the Korean Government and
the Grant No. R01-2005-000-10404-0 from the Basic Research Program of the Korea
Science \& Engineering Foundation (W.S.C., Y.G.K., C.B.P and Y.S.).

\section*{Appendix A.}

The one-loop beta function coefficients $b_a$ and anomalous
dimension $\gamma_i$ in the MSSM are given by 
\begin{eqnarray} 
b_3&=&-3, \qquad
b_2=1,\qquad b_1=\frac{33}{5},
\nonumber \\
\gamma_{H_u} &=& \frac{3}{2}g_2^2+\frac{1}{2}g_Y^2 -3y_t^2,
\nonumber \\
\gamma_{H_d} &=& \frac{3}{2}g_2^2+\frac{1}{2}g_Y^2 - 3 y_b^2 -  y_\tau^2
\nonumber \\
\gamma_{Q_a} &=& \frac{8}{3} g_3^2 + \frac{3}{2} g_2^2
                +\frac{1}{18} g_Y^2 - (y_t^2 + y_b^2) \delta_{3a},
\nonumber \\
\gamma_{U_a} &=& \frac{8}{3} g_3^2  + \frac{8}{9} g_Y^2
                - 2 y_t^2 \delta_{3a},
\nonumber \\
\gamma_{D_a} &=& \frac{8}{3} g_3^2 + \frac{2}{9} g_Y^2
                - 2 y_b^2 \delta_{3a},
\nonumber \\
\gamma_{L_a} &=& \frac{3}{2} g_2^2 + \frac{1}{2} g_Y^2
                - y_\tau^2 \delta_{3a},
\nonumber \\
\gamma_{E_a} &=& 2 g_Y^2 - 2 y_\tau^2 \delta_{3a}, 
\end{eqnarray} 
where $g_2$ and $g_Y=\sqrt{3/5}g_1$ denote the $SU(2)_L$ and $U(1)_Y$
gauge couplings.

The $\theta_i$ and $\dot{\gamma}_i$ which determine the soft scalar
masses at $M_{GUT}$ are given by 
\begin{eqnarray} 
\theta_{H_u} &=&
3g_2^2+g_Y^2 -6y_t^2(a_{H_u}+a_{Q_3}+a_{U_3}), \nonumber \\
\theta_{H_d} &=& 3g_2^2+g_Y^2 - 6y_b^2(a_{H_d}+a_{Q_3}+a_{D_3}) -
2y_\tau^2(a_{H_d}+a_{L_3}+a_{E_3})
\nonumber \\
\theta_{Q_a} &=& \frac{16}{3} g_3^2 + 3 g_2^2
                +\frac{1}{9} g_Y^2 - 2\Big(y_t^2(a_{H_u}+a_{Q_3}+a_{U_3}) + y_b^2(a_{H_d}+a_{Q_3}+a_{D_3})\Big) \delta_{3a},
\nonumber \\
\theta_{U_a} &=& \frac{16}{3} g_3^2  + \frac{16}{9} g_Y^2
                - 4y_t^2(a_{H_u}+a_{Q_3}+a_{U_3}) \delta_{3a},
\nonumber \\
\theta_{D_a} &=& \frac{16}{3} g_3^2 + \frac{4}{9} g_Y^2
                - 4y_b^2(a_{H_d}+a_{Q_3}+a_{D_3}) \delta_{3a},
\nonumber \\
\theta_{L_a} &=& 3 g_2^2 + g_Y^2
                - 2y_\tau^2 (a_{H_d}+a_{L_3}+a_{E_3})\delta_{3a},
\nonumber \\
\theta_{E_a} &=& 4 g_Y^2 - 4 y_\tau^2(a_{H_d}+a_{L_3}+a_{E_3})
\delta_{3a}, 
\end{eqnarray} 
and 
\begin{eqnarray} 
\dot\gamma_{H_u} &=& \frac{3}{2} g_2^4
+ \frac{11}{2} g_Y^4
                     - 3 y_t^2 b_{y_t},
\nonumber \\
\dot\gamma_{H_d} &=& \frac{3}{2} g_2^4 + \frac{11}{2} g_Y^4
                    - 3 y_b^2 b_{y_b} - y_\tau^2 b_{y_\tau},
\nonumber \\
\dot \gamma_{Q_a} &=&  -8 g_3^4 + \frac{3}{2} g_2^4 + \frac{11}{18} g_Y^4
                  -(y_t^2 b_{y_t} + y_b^2  b_{y_b}) \delta_{3a},
\nonumber \\
\dot\gamma_{U_a} &=& - 8 g_3^4  +  \frac{88}{9} g_Y^4
                   - 2 y_t^2 b_{y_t} \delta_{3a},
\nonumber \\
\dot\gamma_{D_a} &=& - 8 g_3^4 + \frac{22}{9} g_Y^4
                     - 2 y_b^2 b_{y_b} \delta_{3a},
\nonumber \\
\dot\gamma_{L_a} &=& \frac{3}{2}g_2^4 + \frac{11}{2} g_Y^4
                     - y_\tau^2 b_{y_\tau} \delta_{3a},
\nonumber \\
\dot\gamma_{E_a} &=& 22 g_Y^4 - 2 y_\tau^2 b_{y_\tau} \delta_{3a},
\end{eqnarray}
where
\begin{eqnarray}
b_{y_t} &=& - \frac{16}{3} g_3^2 - 3 g_2^2 - \frac{13}{9} g_Y^2
             + 6 y_t^2 + y_b^2,
\nonumber \\
b_{y_b} &=& - \frac{16}{3} g_3^2 - 3 g_2^2 - \frac{7}{9} g_Y^2
             + y_t^2 + 6 y_b^2 + y_\tau^2,
\nonumber \\
b_{y_\tau} &=& - 3 g_2^2 - 3 g_Y^2 + 3 y_b^2  + 4 y_\tau^2. 
\end{eqnarray}



\begin{thebibliography}{99}

\bibitem{MSSM} H.P. Nilles, Phys. Rept. {\bf 110} (1984) 1; H.E. Haber and G.L.
   Kane, Phys. Rept. {\bf 117} (1985) 75; S.P. Martin, in Perspectives on
   Supersymmetry, ed. G.L. Kane, pp.1--98 [hep--ph/9709356].


\bibitem{kklt}
S.~Kachru, R.~Kallosh, A.~Linde and S.~P.~Trivedi,
Phys.\ Rev.\ D {\bf 68}, 046005 (2003) [arXiv:hep-th/0301240].

\bibitem{choi1}
  K.~Choi, A.~Falkowski, H.~P.~Nilles, M.~Olechowski and S.~Pokorski,
  JHEP {\bf 0411}, 076 (2004)
  [arXiv:hep-th/0411066];
  K.~Choi, A.~Falkowski, H.~P.~Nilles and M.~Olechowski,
Nucl. Phys. {\bf B718}, 113 (2005) [arXiv:hep-th/0503216].

\bibitem{Choi:2005uz}
  K.~Choi, K.~S.~Jeong and K.~i.~Okumura,
  JHEP {\bf 0509}, 039 (2005)
  [arXiv:hep-ph/0504037].

\bibitem{mirage3}
  M.~Endo, M.~Yamaguchi and K.~Yoshioka,
  Phys.\ Rev.\ D {\bf 72}, 015004 (2005)
  [arXiv:hep-ph/0504036];
  A.~Falkowski, O.~Lebedev and Y.~Mambrini,
  JHEP {\bf 0511}, 034 (2005)
  [arXiv:hep-ph/0507110];
   K.~Choi, K.~S.~Jeong, T.~Kobayashi and K.~i.~Okumura,
  Phys.\ Lett.\ B {\bf 633}, 355 (2006)
  [arXiv:hep-ph/0508029];
  R.~Kitano and Y.~Nomura,
  Phys.\ Lett.\ B {\bf 631}, 58 (2005)
  [arXiv:hep-ph/0509039];
 O. Loaiza-Brito, J. Martin, H.
P. Nilles and  M. Ratz, hep-th/0509158; O.~Lebedev, H.~P.~Nilles and
M.~Ratz,
  arXiv:hep-ph/0511320;
%
R.~ Kitano and Y.~ Nomura,
Phys. Rev. D {\bf 73}, 095004 (2006)
[arXiv:hep-ph/0602096];
%
A.~Pierce and J.~Thaler,
  JHEP {\bf 0609}, 017 (2006)
  [arXiv:hep-ph/0604192];
%
K.~Kawagoe and M.~Nojiri,
Phys. Rev. D {\bf 74}, 115011 (2006)
[arXiv:hep-ph/0606104];
%
  H.~Baer, E.~K.~Park, X.~Tata and T.~T.~Wang,
  JHEP {\bf 0608}, 041 (2006)
  [arXiv:hep-ph/0604253];
%
Phys. Lett. {\bf B 641}, 447 (2006) [arXiv:hep-ph/0607085];
H. Baer, E. K. Park, X. Tata and T. T. Wang, 
hep-ph/0703024.


\bibitem{cklos}
K. Choi, K. Y. Lee, Y. Shimizu,
Y. G. Kim and K.-i. Okumura, JCAP {\bf 0612}, 017 (2006)
[arXiv:hep-ph/0609132].

\bibitem{LHC}ATLAS Collaboration, Technical Design Report, CERN/LHCC/99--15
   (1999); CMS Collaboration, Technical Proposal, CERN/LHCC/94--38 (1994);
   J.G. Branson, D. Denegri, I. Hinchliffe, F. Gianotti, F.E. Paige and
   P. Sphicas [ATLAS and CMS Collaborations], Eur. Phys. J. directC {\bf 4}
   (2002) N1.

\bibitem{beenakker}
W.~Beenakker, R.~Hopker, M.~Spira and P.~M.~Zerwas, 
Nucl. Phys. B {\bf 492}, 51 (1997)
[arXiv:hep-ph/9610490].


\bibitem{Bachacou:1999zb}
  I.~Hinchliffe, F.~E.~Paige, M.~D.~Shapiro, J.~Soderqvist and W.~Yao,
  Phys.\ Rev.\ D {\bf 55}, 5520 (1997)
  [arXiv:hep-ph/9610544];
  H.~Bachacou, I.~Hinchliffe and F.~E.~Paige,
  Phys.\ Rev.\ D {\bf 62} (2000) 015009
  [arXiv:hep-ph/9907518].

\bibitem{weiglein}
G.~Weiglein {\it et al.} [LHC/LC Study Group], Phys. Rept. {\bf 426} (2006) 47
[arXiv:hep-ph/0410364].


\bibitem{wmap}
 WMAP Collaboration,
D.~N.~Spergel {\it et al.} [arXiv:astro-ph/0603449];
Astrophys. J.  Suppl. Ser. {\bf 148} (2003) 175.

\bibitem{pythia}
 T. Sjostrand, P. Eden, C. Friberg, L. Lonnblad, G. Miu, S. Mrenna and  
 E. Norrbin, Computer Physics Commun. 135 (2001) 238;
 T. Sjostrand, S. Mrenna and P. Skands, 
 LU TP 06-13, FERMILAB-PUB-06-052-CD-T [hep-ph/0603175].

\bibitem{PGS}
http://www.physics.ucdavis.edu/~conway/research/software/pgs/pgs4-general.htm

\bibitem{dkn00}
M.~Drees, Y.G.~Kim, M.M.~Nojiri, D.~Toya, K.~Hasuko and T.~Kobayashi,
Phys. Rev. {\bf D 63} (2001) 035008 [arXiv:hep-ph/0007202].


\bibitem{lester00}
C.~G.~Lester and D.~J.~Summers,
Phys. Lett. {\bf B 463} (1999) 99-103.

\end{thebibliography}
\end{document}